%% file: root.tex
\PassOptionsToPackage{pdftex, xetex}{graphicx}
\PassOptionsToPackage{usenames,dvipsnames,svgnames,table}{xcolor}
\documentclass[letterpaper, 10 pt, conference]{ieeeconf}
\IEEEoverridecommandlockouts
\makeatletter
\let\NAT@parse\undefined
\makeatother
\usepackage[sort,numbers,compress]{natbib}

\usepackage{pratik}
\graphicspath{{./figures42/}}

\usepackage[color=gray!30,textsize=tiny]{todonotes}

\makeatletter
\DeclareMathSizes{\@xpt}{9}{7}{5}
\makeatother
\title{A Model for Multi-Agent Heterogeneous Interaction Problems}

\author{Christopher D. Hsu$^{1,2}$, Mulugeta A. Haile$^{1}$, and Pratik Chaudhari$^{2}$% <-this % stops a space
\thanks{$^{1}$DEVCOM Army Research Laboratory
\href{mailto:christopher.d.hsu.civ@army.mil}{christopher.d.hsu.civ@army.mil},
\href{mailto:mulugeta.a.haile.civ@army.mil}{mulugeta.a.haile.civ@army.mil}}
\thanks{$^{2}$Department of
Electrical \&  Systems Engineering and the GRASP
Laboratory at the University of Pennsylvania.
\href{mailto:chsu8@seas.upenn.edu}{chsu8@seas.upenn.edu},
\href{mailto:pratikac@seas.upenn.edu}{pratikac@seas.upenn.edu}}
\thanks{Code: \href{https://github.com/grasp-lyrl/Model4MAInteractions}{https://github.com/grasp-lyrl/Model4MAInteractions}}
}

\begin{document}

% \listoftodos
% \clearpage

\maketitle
\thispagestyle{empty}

\begin{abstract}
We introduce a model for multi-agent interaction problems to understand how a heterogeneous team of agents should organize its resources to tackle a heterogeneous team of attackers. This model is inspired by how the human immune system tackles a diverse set of pathogens. The key property of this model is a ``cross-reactivity'' kernel which enables a particular defender type to respond strongly to some attacker types but weakly to a few different types of attackers. We show how due to such cross-reactivity, the defender team can optimally counteract a heterogeneous attacker team using very few types of defender agents, and thereby minimize its resources. We study this model in different settings to characterize a set of guiding principles for control problems with heterogeneous teams of agents, e.g., sensitivity of the harm to sub-optimal defender distributions, and competition between defenders gives near-optimal behavior using decentralized computation of the control. We also compare this model with existing approaches including reinforcement-learned policies, perimeter defense, and coverage control.
% teams consisting of a small number of attackers and defenders, estimating and tackling an evolving attacker distribution, and 
\end{abstract}

% \clearpage
% \input{reviews}
% \clearpage

% \input{presentation}
\input{intro}
\input{related}
\input{problem}
\input{implementation}

\input{basic}

\input{competition}
\input{RL}
\input{experiments}

\input{conclusion}
% \input{mobile}

\begin{footnotesize}
\bibliographystyle{ieeetr}
\bibliography{references}
\end{footnotesize}

% \input{appendix}
% \clearpage
% \input{limits}
% \input{kf}

\end{document}

%% file: intro.tex
% !TEX root = ./root.tex

\section{Introduction}
\label{s:intro}
Consider the adaptive immune system which protects an organism from pathogens. Some pathogens are persistent but benign (e.g, common cold) and others are rare but dangerous (e.g., HIV). There is a vast number ($\sim 10^{30}$) of other pathogens on this spectrum. Tackling every pathogen optimally requires a specialized receptor, lymphocytes generate these receptor proteins which bind to the pathogens. The immune system has evolved to achieve a somewhat counter-intuitive solution: it allocates a relatively larger amount of resources to tackling pathogens that cause a large harm even if they are rare, and relatively fewer resources to pathogens which cause less harm, even if they are encountered frequently. Mathematical models of the immune system such as~\cite{Mayer_organized} have argued that two very natural principles are sufficient to explain this phenomenon. First, the metabolic resources spent by an organism are proportional to the total number of diverse receptors (as opposed to the number of receptors). Therefore, the immune repertoire minimizes the diversity of the receptors. Second, even if a receptor is best suited to tackling a specific pathogen, it can also tackle other pathogens with similar protein structures with a small probability. Therefore, the repertoire can tackle a much more heterogeneous pathogen population than simply what the composition of the receptors suggests, if it has the right composition.

\begin{wrapfigure}{r}{0.5\linewidth}
\centering
\includegraphics[width=\linewidth]{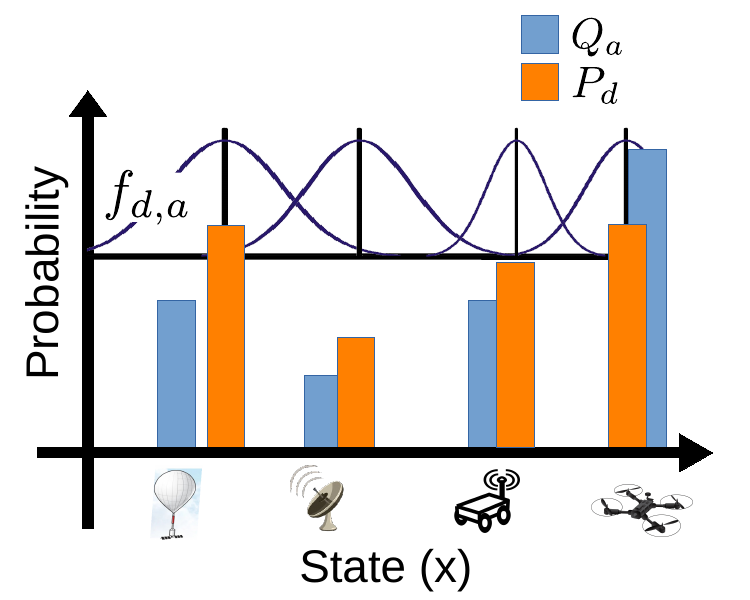}
\caption{Orange defenders from distribution $P_d$ successfully interact with blue attackers from distribution $Q_a$ with probability $f_{d,a}$ which depends on the defender type $d$ and the attacker type $a$.}
\label{fig:fig1}
\end{wrapfigure}
The goal of this paper is to demonstrate that multi-agent interaction problems, where a team of defenders interacts with a team of attackers, also benefit from these principles. \cref{fig:fig1} shows a scenario with different electronic, ground, and air systems as attackers and defenders. Each team consists of a heterogeneous group of agents with different types or capabilities, say offensive vs. defensive skills, sensing modalities, operating environments, and/or mobility. The blue/orange bars show the relative composition of different types in the attacker/defender teams, e.g., in the picture, the attackers have a large number of aerial vehicles. The problem that we will focus upon is as follows.
\begin{quote}
\textbf{Problem:} Given a particular composition of the attacker team, i.e., the relative proportion of the \emph{types} of agents, what should the composition of the defender team be in order to minimize harm.
\end{quote}
As we discuss in~\cref{s:related_work} and via examples in the rest of the paper, applications of this work range from pursuit-evasion games, perimeter defense and coverage control problems for multiple agents, to optimal resource allocation for teams in the face of diverse and conflicting objectives.

\subsection{Contributions}

\textbf{We use a mathematical model (\cref{s:problem}) of the human adaptive immune system first proposed in~\cite{Mayer_organized} to understand how a defending team can optimally allocate its resources to minimize the harm incurred from a heterogeneous team of attackers.} We focus our analysis on two situations in~\cref{s:basic}: (i) when no single type of defender can defend against every type of attacker, and (ii) when defenders have limited resources that they should devote optimally to tackle attackers.  
% We also discuss how the defender team can estimate and adapt its composition to an unknown (and evolving) heterogeneous attacker team's composition using information obtained from their encounters.

\textbf{In~\cref{s:competition} we show that in this model, a decentralized control policy where defender agents compete with each other for successful encounters with attacking agents can achieve near-optimal harm even when the attacking distribution is not known.}
Centralized computation may seem necessary to select the optimal defender distribution because it depends upon both local and global structures of the attacker team's composition. We study the ``centralized estimation and decentralized control'' setting where information obtained from individual interactions with the attackers is shared with all the defenders but different defenders use this information in different ways, e.g., deciding which attacker type to tackle. This competition between defenders leads to the proliferation of successful defender types over multiple episodes.

\textbf{We compare and contrast our model and control policies with three types of existing approaches: multi-agent reinforcement learning (RL) in~\cref{s:RL}, perimeter defense~\cite{Shishika2018} and coverage control~\cite{cortes2004} in~\cref{s:mapping_to_euclidean}.} 
Different choices of the ingredients of our model allow us to study problems like perimeter defense and coverage control which provides a different perspective on them. For example, we show how coverage control can be achieved while minimizing the number of sensors given the capability.

%\subsection{Organization} We formulate the model in~\cref{s:problem}. We investigate the model under different scenarios using theory and simulation experiments in~\cref{s:basic,s:limits,s:KF,s:competition}. Related work is discussed in~\cref{s:related_work}.

%% file: related.tex
% !TEX root = ./root.tex

\section{Related Work}
\label{s:related_work}

Interactive multi-agent problems, e.g., pursuit-evasion games, have seen a wide range of perspectives~\cite{Karaman2010, isaacs1954}.
% and surveys~\cite{Volkan-PEsurvey,robin:hal-01183372}. 
% Problems with multiple agents are complex due to their high dimensional state-space~\cite{Hilal2013AnIS}. 
We apply our model to two variants of interactive multi-agent problems: perimeter defense and coverage control. A large body of the work on perimeter defense~\cite{Lee-GroundvsAir} studies how multiple defenders decompose the problem into smaller games~\cite{Shishika2018}, or reduce the defense strategy to an assignment problem~\cite{Chen-ReachAvoid}. 
Similarly, coverage control~\cite{cortes2004} solves the locational optimization problem for spatial resource-allocation problem and works since improve decentralization~\cite{Schwager2009}, minimize assumptions on prior knowledge on the environment~\cite{luo2018AdaptiveSampling}, and heterogeneity~\cite{Reily2021}.

% We investigate a variant of the pursuit-evasion game first introduced as the target-guarding problem where the pursuer/defender tries to prevent the evader/intruder from reaching the target~\cite{Garcia-coopdeffrommissiles,Shishika_ReviewPDGames}. 
%
%Classical control has been extended to the multi-agent setting~\cite{fierro-MRCoordination, Sukhatme2008} with some applications such as target tracking/sensing problem~\cite{Salam-adaptivesampling} and localization~\cite{Spletzer-cooplocalization}. A wide array of works have been able to incorporate information-based control as their objectives~\cite{Atanasov-DecAIA,Schlotfeldt-ResilientAIA22,Kantaros-scalableAIA,Tzes-distAIG21,Jeong-RLAIA} for better situational awareness~\cite{Salam-sitawareness}.
%
% A theme that has emerged recently in reinforcement learning-based multi-agent control~\cite{Tan93multi-agentreinforcement, lowe2020multiagent} is to obtain decentralized, cooperative policies~\cite{li2021deep} for non-stationary problems with limited information. 
% A popular paradigm is to train a centralized policy and execute it in a decentralized fashion~\cite{son2019qtran}. 
A popular theme in reinforcement learning-based multi-agent control is to train a centralized policy and execute it in a decentralized fashion~\cite{lowe2020multiagent}.
% Emergence of coordination has also been studied~\cite{baker2020emergent}. 
These methods typically suffer from poor sample complexity but some works have scaled them to large problems~\cite{long2020evolutionary}, including problems with up to 1000 attackers and defenders using heuristics~\cite{hsu2020scalable}. There are also bio-inspired approaches mimicking the behavior of ants, bees and flocks~\cite{Hsieh2008BiologicallyIR,Zavlanos-controlforflocking} for multi-agent control.

In the context of the above literature, the place of the present paper is to study a theoretical model where large heterogeneous multi-agent interaction problems can be analyzed precisely. 
% This model has several benefits, e.g., using the model, we can estimate and adapt to an unknown attacker distribution and find that even if the defenders do not accurately estimate the attacker distribution, the eventual harm can be near-optimal. 
This model establishes guiding principles for building multi-agent control policies, e.g., inducing competition among the defender agents for successful encounters can lead to near-optimal harm. 
Also, we can simulate the model even when the number of agents is extremely large. 
A large number of papers, including many above, have sought to understand trends in learned policies and costs for multi-agent systems using simulations. The utility of our model is that it is a more straightforward approach to achieving the same insights.

%This work demonstrates scalable decentralized policies that can adapt to a wide range of changing attacker distributions for problems as large as 1000 attackers and defenders. We also obtain a useful guiding principle: competition among the defenders where successful interaction with the attackers acts as the reward may achieve optimality of the resources spent. Such ideas can be used for designing new reinforcement learning methods for multi-agent control. Prior work has noticed the benefits of a similar competition arising out of stochastic policies.

%Our work is inspired from how the adaptive immune system in humans is organized. We build upon recent literature in biophysics which models the response of the adaptive immune system in an environment with a large number of evolving pathogens~\cite{MayerRemembers,Mayer8630,Mayer5914}. We believe that the salient traits exhibited in the immune system such as heterogeneous defender types with wide cross-reactivity that interact with a range of attackers, and decentralized estimation and execution are key building blocks for building and understanding multi-agent autonomy.

\emph{Modeling} heterogeneity, which is the focus of this paper, has gained interest~\cite{Bettini-HetGPPO, Mitra-HeterogeneityBandit}. Heterogeneity comes in many forms, e.g., differences in roles~\cite{Wang-RODE}, robotic capabilities and or sensors~\cite{Salam-heterosensors}, dynamics~\cite{Edwards-heterocontrol}, and even teams of air and ground robots~\cite{Grocholsky-coopairground}. Heterogeneity has been defined~\cite{TwuEgerstedt2014} for systems in the finite and discrete setting~\cite{Abbas2011DistributionOA,Bezzo2013}, but algorithmic methods that can tackle large-scale heterogeneity have been difficult to build. 
% A taxonomy for heterogeneous multi-robot systems in introduced~\cite{Bettini-HetGPPO} and a method for more explicit handling of the heterogeneous systems is developed for RL. 
Task assignment with heterogeneous agents~\cite{chernova-Strata} is another similar problem to ours, but a desired trait distribution is necessitated by the objective instead of calculating it explicitly. In comparison, the present paper uses a simple formulation to understand what distribution is best and how to allocate heterogeneous agents optimally in a multi-agent interaction problem.

%% file: problem.tex
% !TEX root = ./root.tex

\section{Problem formulation}
\label{s:problem}

\subsection{The model for interactions between attackers and defenders}

\begin{table}[!htbp]
\centering
\footnotesize
\caption{The following key quantities of model are collected here for reference; they are introduced in the text.}
\renewcommand{\arraystretch}{1.25}
\begin{tabular}{cc}
\toprule
$Q_a$ & Probability of attacker of type $a$ (attacker distribution)\\
$P_d$ & Probability of defender of type $d$ (defender distribution)\\
$f_{d,a}$ & Probability of recognition of type $a$ by type $d$ (cross-reactivity)\\
$\tilde P_a$ & Coverage (function of $P_d$ and $f_{d,a}$)\\
$\bar F_a$ & Harm from attacker of type $a$ \\
$N_a$ & Number of attackers of type $a$ (used for simulation)\\
$N_d$ & Number of defenders of type $d$ (used for simulation)\\
% $D$   & Type discretization width (used for simulation)\\
$N$   & Number of types/intervals in the shape space $x$\\
\bottomrule
\end{tabular}
\end{table}

\paragraph{Shape/State space} In biology, attackers of type $a$ and defenders of type $d$ interact in an abstract space (often also a metric space) called the shape-space~\cite{PERELSON1979645}. For the purposes of this paper, we think of the shape space $x$ as the real-valued space of the different types of agents; if attacker types $a, a'$ are similar to each other then $\abs{a - a'}$ is small. The mathematical formulation that follows will be general and can also be used for shape-spaces that are multi-dimensional or discretized. We will show several examples where the mathematical formulation can be directly mapped to classical problems (e.g., perimeter defense, coverage control, etc.) by thinking of the shape space as the Euclidean locations of the attacker and defender agents.

\paragraph{Composition of the team} Let $Q_a$ be the probability that denotes the next attack will be caused by an attacker of type $a$. Let $P_d$ be the probability of a defender of type $d$ being present. The composition of the attacker and defender teams under consideration is the number of distinct types that $Q_a$ and $P_d$ respectively put their probability mass on. In the mathematical model, we do not explicitly model the number of attacker and defender agents explicitly, effectively we imagine an infinite number of agents sampled according to their respective distributions $Q_a$ and $P_d$ interacting with each other. In the numerical experiments, we will indeed have a finite number of agents for both sides independent of the number of types.

\paragraph{Interactions between attackers and defenders of different types}
We model the interaction between an attacker of type $a$ with a defender as a Poisson random variable with rate $\l_a(t)$. In biology, the longer an attacker type exists, the larger the number of agents of that type, and therefore higher the probability that the particular pathogen type interacts with a lymphocyte receptor. We can mimic this as a rate of interaction that increases exponentially with time, i.e., $\dot{\l}_a(t) = \l_a \nu'_a$ starting from some initial rate $\l_a(0) = 1$. For our problem, there will be situations where a particular type of attacker reinforces the team composition by having more agents of its type. Under this assumption, the expected number of interactions of some defender with an attacker of type $a$ is $m_a(t) = \int_0^t \dd{\t} \l_a(\t) \approx \l_a(0) (e^{\nu'_a t} - 1)/\nu'_a$.

In biology, a successful interaction means that the immune repertoire has produced the correct receptor $d$ to tackle pathogens of type $a$; this is called a ``recognition'' and it occurs, say, within a few days of the infection. Unsuccessful interactions of defenders with the attackers in the meanwhile incur harm to the organism. We will model recognition as a probabilistic event. Let the ``cross-reactivity'' $f_{d,a}$ denote the probability that defender of type $d$ successfully interacts with an attacker of type $a$. There are different models for the interaction of different types of receptors/pathogens and consequently for different types of defenders/attacker agents; we use a Gaussian
\[
    f_{d,a} \propto \exp\rbr{-(d-a)^2/(2\s^2)}
\]
where $\s$ denotes the bandwidth. Defenders of type $d$ that have a low probability under this probability distribution have a small chance of recognizing an attacker of type $a$. The essential conclusions of our mathematical formulation do not change if the cross-reactivity is different. Using a Gaussian simply ensures that we can perform some of the calculations analytically, e.g., compute the optimal defender team distribution $P_d^*$. The cross-reactivity being Gaussian is not essential for our numerical simulations.

We will assume that an attacker type does not cause any harm after a recognition event. In biology, this is because there are appropriate receptors to tackle it. In our problem, this is because we imagine that the appropriate defender type can neutralize all agents of the recognized attacker type easily. This is a modeling choice and this way we can focus on computing the optimal defender composition instead of modeling the number of agents.

If the defender distribution is $P_d$, then the total probability of recognizing an attacker of type $a$ is $\tilde P_a = \sum_d f_{d,a} P_d$. We will call this the ``coverage''. A naive defender composition would maximize coverage but that would not take into account the harm caused by each attacker type.

\paragraph{Harm caused by an attacker type}
Let $F_a(t)$ be the harm caused by an attacker of type $a$ until it is recognized at time $t$. We will assume that this grows exponentially $\dot{F}_a(t) = F_a \nu_a$ starting from some $F_a(0) = 1$. The exponents $\nu_a$ here and $\nu_a'$ in the expression for the expected number of interactions of attacker $a$ denoted by $m_a(t)$ above can be different. This is because the harm incurred (say, proportional to the actual number of attackers $a$) can be different than the number of attackers $a$ observed by the defenders. For large times $t \approx \log(m_a/\l_a(0))/\nu'_a$, the expected harm caused by an attacker of type $a$ after $m$ interactions is
\beq{
    F_a(m) = F_a(0) \rbr{\f{m}{\l_a(0)}}^{\nu_a/\nu'_a} \propto m^\a;
    \label{eq:fm}
}
where $\a = \nu_a/\nu'_a$~\cite{Mayer_organized}. Observe that in this case, $F_a(m)$ is polynomial in the number of interactions $m$. In this paper we will assume unsuccessful interactions of defenders with attackers cause one unit of harm, e.g. $\alpha=1$, $F_a(m) = m$, and $\Dot{m}_a(t) = 1$. The harm function can also take other forms, e.g., $F_a(m) = 1 - e^{-\b m}$ would model the situation where the harm plateaus after a large number of interactions. The harm $\bar F_a$ caused by an attacker of type $a$ until it is recognized by the defenders is thus
\beq{
    \Bar{F}_a(P_d) = \Tilde{P}_a \int_0^{\infty} \dd{m} F_a(m) e^{-m \Tilde{P}_a}.
    \label{eq:Fbar}
}
We call this quantity the ``empirical harm'' because we will be able to estimate it by simulating interactions between attackers of type $a$ sampled from $Q_a$ and defenders of type $d$ sampled from $P_d$. The harm $\bar F_a$ is a function of the defender distribution $P_d$ through the coverage $\tilde P_a$.

In the above formulation, we have avoided modeling the \emph{number} of agents in the attacker and defender teams and used quantities like $m_a(t)$ (the expected number of interactions of an attacker type $a$) and $F_a(t)$ (the harm caused by $a$ if it is not recognized by time $t$) to implicitly capture the number of agents. This choice simplifies analytical calculations and it is conceptually equivalent to considering an infinite number of attackers and defenders that interact probabilistically. Our numerical simulations are conducted with a finite number of agents which we show to be consistent with those from the analytical model.

\paragraph{Minimizing the harm}
Our goal is to minimize
\beq{
    \text{Harm}(P_d) = \textstyle \sum_a Q_a \Bar{F}_a.
    \label{eq:harm}
}
caused by the attacker team with distribution of types given by $Q_a$; note that it is a function of the defender distribution $P_d$ through $\Bar{F}_a$. When $F_a(m) = m^\a$, we can show that the optimal value of this objective is achieved for
\beq{
    \bar F_a = \G(1+\alpha)/\tilde{P}_a^\alpha,
    \label{eq:Fbar*}
}
where $\G$ is the Gamma function~\cite{Mayer_organized}. We will refer to the corresponding harm as the ``analytical harm'' in the rest of the paper. The harm incurred for different settings, e.g., a sub-optimal defender distribution $P_d$, a reinforcement learned-policy, etc. can all be compared to this quantity.

%% file: implementation.tex
% !TEX root = ./root.tex

\subsection{Numerical simulations of the model}
\label{s:numerical_sim}

For numerical implementation, we discretize the shape-space $x \in [0,1]$ into $N$ intervals of equal widths $1/N$ and therefore $a, d \in \cbr{1/N, 2/N, \ldots, 1}$. We first solve the optimization problem in~\cref{eq:harm} to minimize the harm caused by attacker types with distribution $Q_a$ with respect to defender types with distribution $P_d$, using sequential least squares programming (SLSQP) with constraints
\beq{
    \aed{
    \min_{P_d} \quad & \sum_a Q_a \Gamma(1+\alpha)/(\sum_d f_{d,a}P_d)^{\a}\\
    \text{such that}\quad  & P_d \geq 0 \text{ and } \sum_d P_d = 1,
    }
    \label{eq:numerical_harm}
}
where the constraints enforce that $P_d$ is a probability distribution over the shape space. We have used the analytical expression for $\bar F_a$ from~\cref{eq:Fbar*}.

\paragraph{Simulating interaction episodes between attackers and defenders}
The episode begins by sampling a total $\sum_a N_a$ (typically $\sim 100$) attackers from the multinomial distribution $Q_a$; note that some attacker types $a$ may not take part in the simulation episode even if $Q_a > 0$. We are interested in calculating the empirical harm~\cref{eq:Fbar} of the defender team $P_d^*$ obtained by solving~\cref{eq:numerical_harm}. At each time-step, for each attacker type in the episode, we draw a Poisson random variable according to rate $\l_a(t)$ which reflects the interaction of that type with some defender. Each attacker agent of type $a$ interacts with a random defender agent of type, say $d$; in some experiments, they will interact greedily with the closest defender in shape space. For each interaction, we sample a Bernoulli random variable according to the cross-reactivity kernel $f_{d,a}$ that determines whether this interaction is successful or not. If the defender recognizes the attacker, then we set the attacker type's growth rate $\l_a(t)$ to zero (and all agents of that type do not play a further role in the episode). If not, we incur a harm equal to $F_a(m_a(t))$; note that this is different for each attacker type $a$ depending upon their total previous interactions $m_a(t)$. The growth rate for the next time-step is updated using $\dot \l_a(t) = \l_a \nu_a'$ in this case. The simulation ends when all attacker types are successfully recognized by the defenders, i.e., when $\sum_a \lambda_a(t) = 0$.

We implemented this model using vectorized operations and can simulate hundreds of episodes and hundreds of attacker and defender agents. We can also modify the details of the simulation easily to consider situations when there is a finite number of defenders, changing attacker or defender distributions during, and across episodes.

%% file: basic.tex
\section{The optimal defender team has a finite number of defender types}
\label{s:basic}

\subsection{Gaussian attacker and defender distributions}

\begin{figure}[!htpb]
\centering
\includegraphics[width=0.44\linewidth]{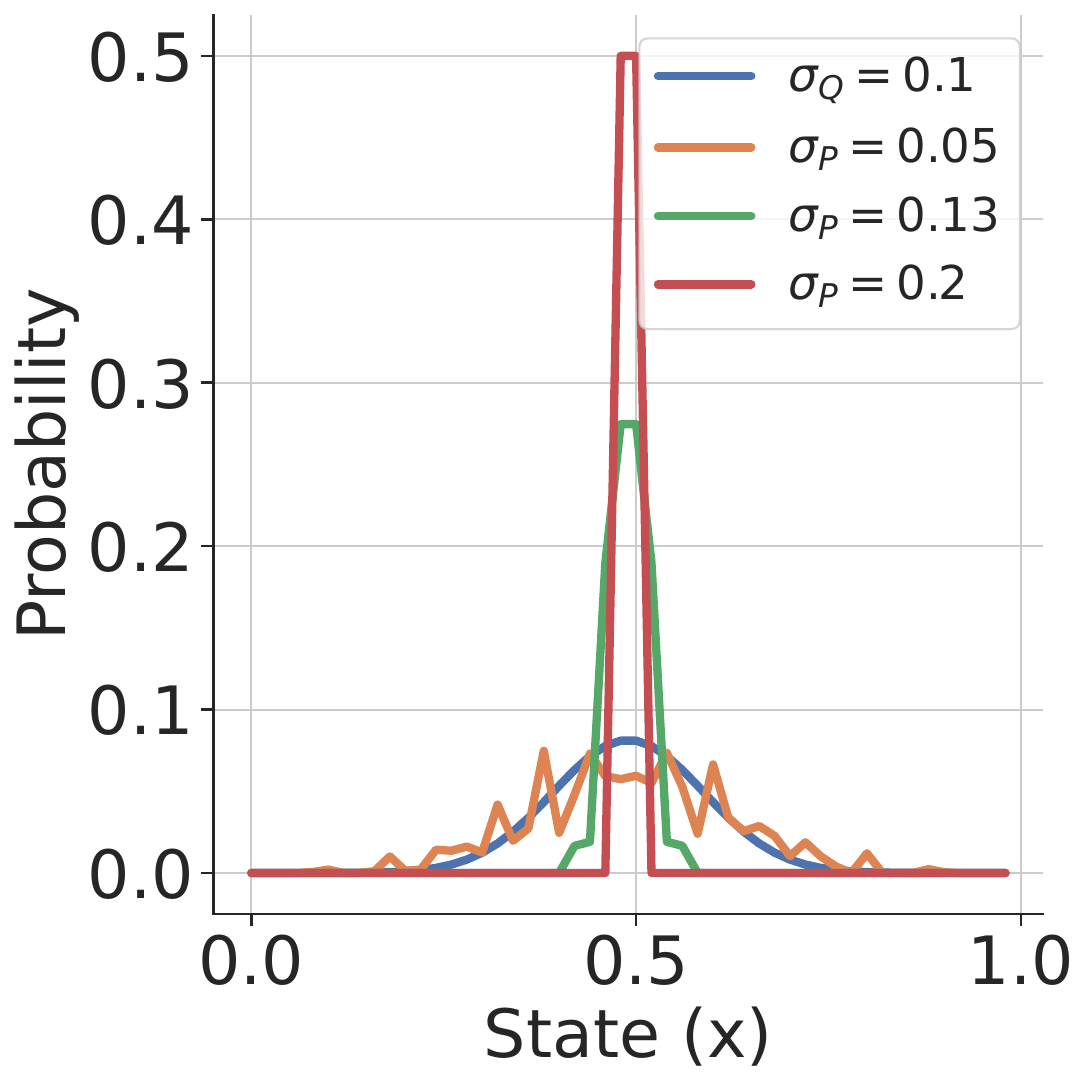}
\includegraphics[width=0.44\linewidth]{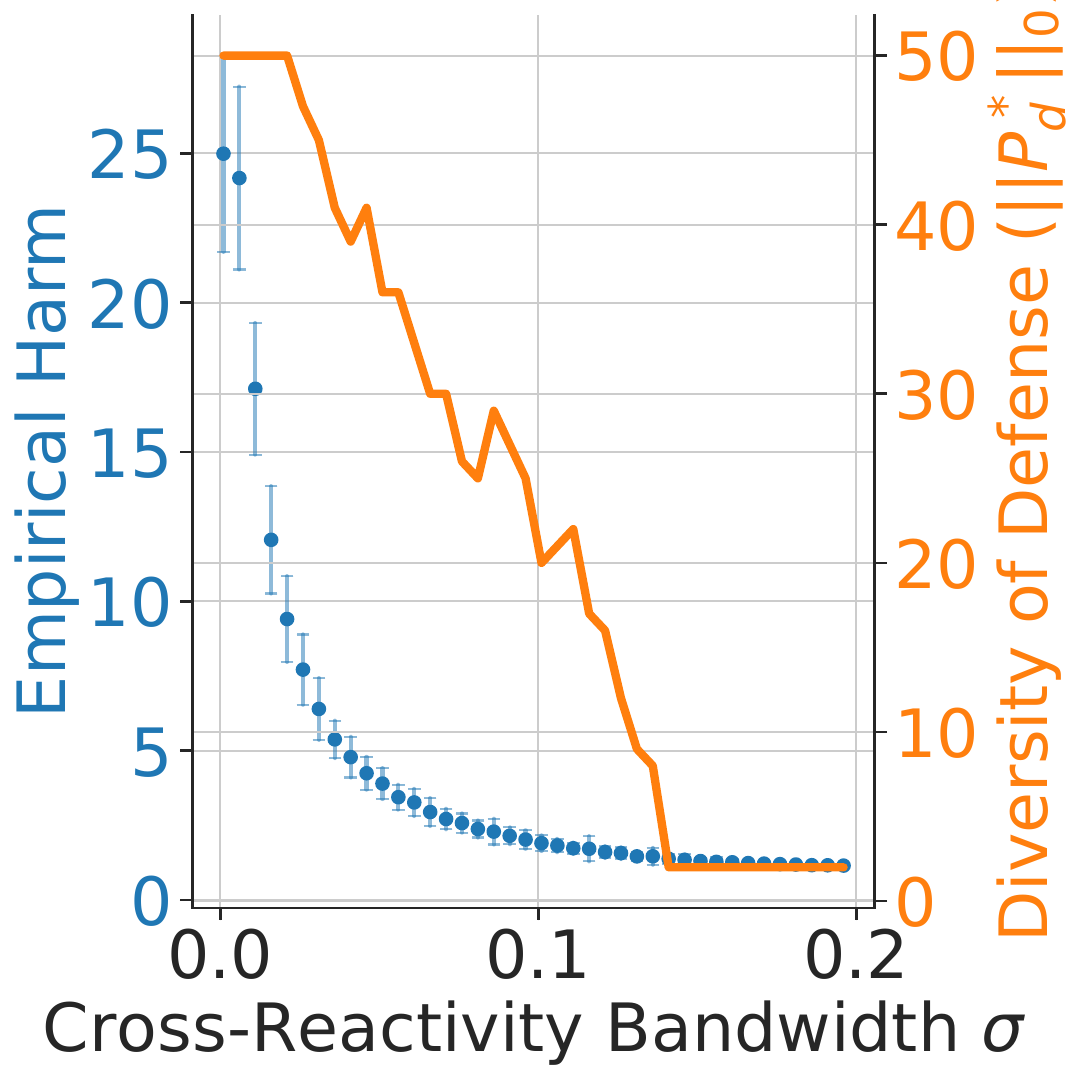}
\caption{A simulation of $\sum_a N_a=100$ attackers sampled from a Gaussian $Q_a$ with $\s_Q=0.1$ interacts with $\sum_d N_d=100$ defenders sampled from $P_d^*$ for different values of $\s_P$ in a shape space $x \in [0,1]$ with $N=50$ types ($\Delta x = 0.02$). \textbf{Left:} For $\a=1$, when $\s_P \geq \s_Q\sqrt{2}$ the optimal $P_d^*$ which is a Gaussian tends towards a Dirac delta distribution at the origin. \textbf{Right: } In the simulation, as we increase $\s_P$ beyond $\s_Q \sqrt{2}$, the empirical harm, i.e., the average unsuccessful interactions until all attackers are recognized, decreases (blue). The number of distinct defender types also decreases (orange).
}
\label{fig:gauss}
\end{figure}

For a Gaussian $Q_a \propto e^{-a^2/(2 \s_Q^2)}$, it can be shown that when a chosen $\s$ is smaller than a threshold $\s_Q \sqrt{1+\a}$, the optimal defender distribution is also Gaussian~\cite{Mayer_organized}
\[
    P_d^* \propto e^{-\f{d^2}{2 ( (1 + \a)\s_Q^2 -\s^2 )}}.
\]
The variance is negative if $\s/\s_Q > \sqrt{1+\a}$. So if the chosen bandwidth is above this threshold, the optimal defense is a Dirac delta at the origin. This shows that larger the cross-reactivity $\s$ it creates concentrated distributions and reduces the number of defender types and diversity of the problem in the finite domain. \cref{fig:gauss} shows a numerical simulation. We can also see this by turning the variational problem in~\cref{eq:harm} into a standard optimization problem by assuming a parametric form for the defender distribution $P_d \propto e^{-d^2/(2 \s_P^2)}$ (by symmetry it has to be centered at $d=0$). For a Gaussian $Q_a$, the solution that minimizes the harm in~\cref{eq:harm} has $\s_P = 0$. In biology, cross-reactivity allows a particular type of receptor to bind to different types of pathogens to varying degrees; a non-zero bandwidth $\s$ thus reduces the number of distinct defender types. It suggests that for some multi-agent interaction problems, the optimal defender composition need not span the entire spectrum of attacker types.

Even if a defender distribution $P_d^*$ is supported on a discrete set in the shape space and has a finite set of defender types, it can tackle many different attacker types with nonzero cross-reactivity. We can therefore design the defender team with fewer resources (types of agents) and still tackle diverse attacker types. Note that it is minimal in terms of types, not in terms of numbers; our model is not suited to addressing the optimal \emph{number} of defenders. Larger the cross-reactivity bandwidth $\s$, smaller the number of types in the optimal defender team $P_d^*$\footnote{In experiments, we measure the diversity of the defender types using the $\ell_0$ norm, smaller the value of $\sum_d \mathbf{1}\{P_d^* > 0\}$ smaller the number of defender types in the optimal defender distribution.}. This is consistent with the intuition that if each defender type can tackle a range of attacker types, even if some of them sub-optimally, we can build a defender team with fewer types. But the unique quantitative insight provided by this model is that if defenders agents are generalists beyond a specific threshold ($\s_P > \s_Q \sqrt{1 + \a}$ in the Gaussian example above), then having only a few types is \emph{optimal}.

% Let us note that such discrete solutions are seen in many other problems, e.g., capacity of a Gaussian channel under an amplitude constraint~\cite{smith1971information}, discrete priors in Bayesian statistics~\cite{berger2009formal}, etc.

\subsection{Non-Gaussian attacker and defender distributions}

\begin{figure}
    \centering
    \includegraphics[width=0.44\linewidth]{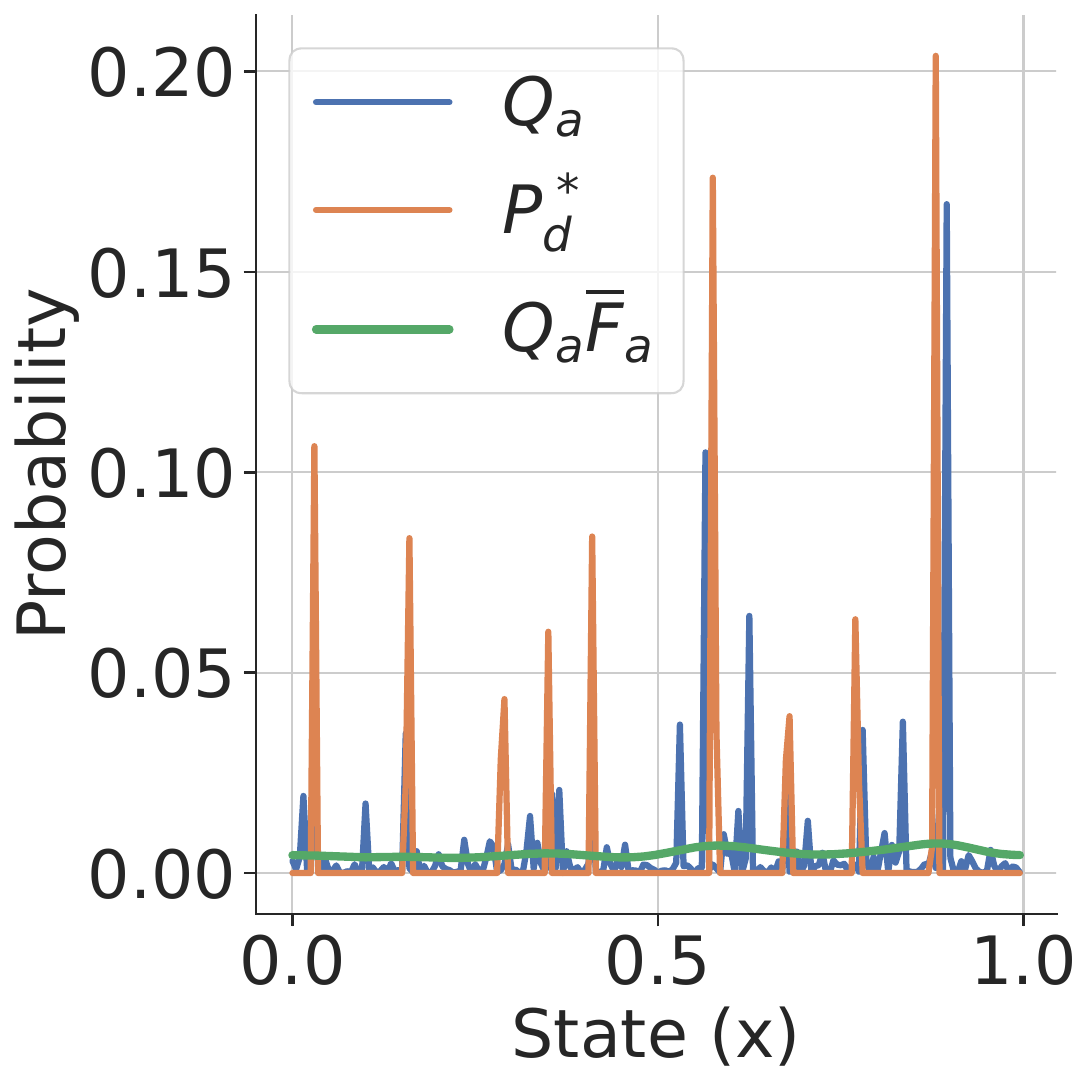}%\\
    \includegraphics[width=0.44\linewidth]{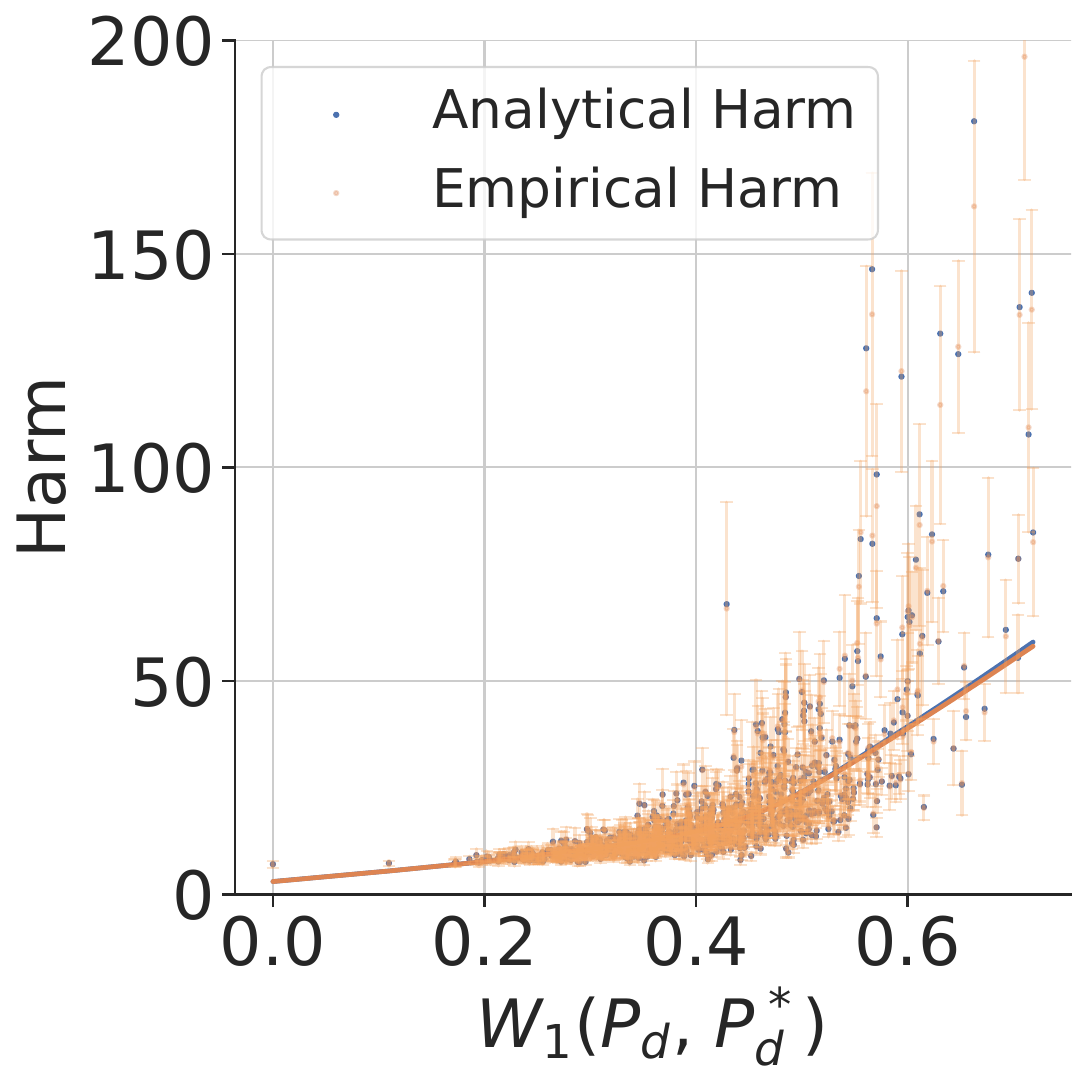}%
    \caption{\textbf{Left:} Simulation of the interaction of defenders (orange) with attackers from distribution $Q_a$ (blue). The optimal defender distribution $P_d^*$ (orange) is found by optimizing~\cref{eq:harm}. Cross-reactivity $f_{d,a}$ with bandwidth $\s = 0.05$ in a state $x$ with $N=200$ types ($\Delta x = 0.005$) leads to a discrete distribution. The harm $Q_a \bar F_a$ caused by attackers of different types (green) is uniform across the domain.
    \textbf{Right:} The harm incurred using a non-optimal $P_d$ increases as the difference measured by the Wasserstein distance between the probability $P_d$ and probability $P_d^*$ increases. To obtain this plot,  we sampled 1000 different $P_d$s (by perturbing the optimal $P_d^*$ using log-normal noise) and computed the empirical and analytical harm against a fixed $Q_a$. This also indicates that the analytical harm~\cref{eq:Fbar*} is close to the mean of the empirical harm over 100 episodes of our experiments using~\cref{eq:Fbar} for a broad regime.
    %
    %\textbf{Right:} Defenders (orange) with cross reactivity bandwidths $\s \in \cbr{0.05,0.01,0.001}$, (large to small bandwidth from left to right) for the domain $\in [0,1]$ split into thirds, interact with attackers from a distribution $Q_a$ (blue). The harm caused by attackers ($Q_a \bar F_a$) on the left (when $\s=0.05$) is small compared to the right third with $\s=0.001$.
    }
    \label{fig:case0}
\end{figure}

In this section, we show how the result above is expected to hold in general. To build a more complicated attacker distribution $Q_a$, for each $a \in \{1/N, 2/N \ldots, 1\}$ with $N=200$ types, we drew a sample from a log-normal distribution with the coefficient of variation $\kappa^2 = \exp(\s_Q^2)-1$ with $\kappa=5$. The value $Q_a$ is then the probability of this sample under the log-normal. As we see in~\cref{fig:case0} this creates an attacker distribution supported on types in different parts of the shape space. We simulate the model numerically for this $Q_a$ as discussed in~\cref{s:numerical_sim}.

\cref{fig:case0} (left) shows the attacker distribution $Q_a$ (blue), the optimal defender distribution $P_d^*$ (orange) and the normalized harm per attacker type $Q_a \Bar{F}_a / \sum_a Q_a \Bar{F}_a$ (green). As expected from the previous section, the defender team obtained from numerical optimization of~\cref{eq:harm} seems to be supported on a few different types. Broadly, these types are similar to those of $Q_a$ (where the cross-reactivity is high), although they do not match exactly. This will be a general theme, the optimal defender types need not simply tackle the attacker types that are most probable. Cross-reactivity allows defender types to be slightly different than attacker types and yet minimize harm. Each attacker type causes roughly the same normalized empirical harm (in green) which is relatively constant across the domain in spite of the discrete-like distribution of the defender distribution. This is an important property of our model. It can be used as a desideratum, or also an evaluation metric, for multi-agent control policies obtained by different methods such as reinforcement learning (see~\cref{s:RL}).

\cref{fig:case0} (right) shows the analytical and empirical harm for sub-optimal $P_d$. We will use the 1-Wasserstein metric~\cite{santambrogioOptimalTransportApplied2015} to measure distances between different defender distributions. This is reasonable because from the previous paragraph, we expect that two distributions $P_d$ with atoms close to each other may incur similar harm, even if the atoms are not exactly the same. The 1-Wasserstein metric for probability distributions on one-dimensional domains is
\(
    W_1(P_d, P_d^*) = \int_0^1 \abs{F_{P_d}(x) - F_{P_d^*}(x)} \dd{x},
\)
where $F_P(x) = \int_0^x P(y) \dd{y}$ is the cumulative distribution function; in our simulations we use a discretization of our shape-space $x \in [0,1]$ and can calculate this integral easily. Our simulation results show that the empirical harm closely tracks the analytical harm in~\cref{fig:case0} (right). Secondly, it show that the model holds when the number of agents is finite and the shape space is discretized. Furthermore, it is noticeable that the variance of the empirical harm across multiple simulations increases as $P_d$ becomes more sub-optimal. This is not directly implied by~\cref{eq:Fbar*} (which only optimizes the average harm). But it is intuitive: as $P_d$ becomes more sub-optimal, the absence of certain defender types in the team leads to proliferation of certain attacker types, until the small cross-reactivity $f_{d,a}$ from some far away defender type manages to recognize the proliferated attacker type. This has an important practical implication. Our model suggests that if the defender team does not have enough different types to build the optimal composition $P_d$, then it should not only minimize the expected harm as is more commonly done in multi-agent control research, but also the variance of the harm, e.g., using the conditional value at risk~\cite{rockafellar2000optimization}.

%% file: competition.tex
\section{Competition between defenders leads to optimal resource allocation}
\label{s:competition}

The optimal defender distributions $P_d^*$ so far were calculated in a centralized fashion. We next discuss how the optimal defender distribution can also emerge from a decentralized computation. The key idea is to exploit cross-reactivity between the defender types, i.e., the fact that multiple defender types can tackle the same attacker type, and set up competition among the defender types to earn the reward of successfully recognizing an attacker type. Biology does things in a similar way: receptors that successfully recognize antigens proliferate at the cost of other receptors~\cite{DEBOER1994375}.

Let $N_d$ be the finite number of defenders of a particular discretized type $d$. We can set
\beq{
    \dv{N_d}{t} = N_d \sbr{\sum_a Q_a \varphi \rbr{\sum_{d'} N_{d'} f_{d',a}} f_{d,a} - c}.
    \label{eq:competition}
}
Here, the constant $c$ is the rate at which defender types $d$ are decommissioned. In the first term, the total interaction of defender type $d$ with different attacker types $\sum_a Q_a f_{d,a}$ is weighted by $\varphi(\cdot)$ which is a decreasing function of its argument. The quantity $\sum_{d'} N_{d'} f_{d',a}$ determines how many other defender types can tackle $a$. If this is large, then the incremental utility of having the specific defender type $d$ diminishes, and its growth rate $\dot{N}_d$ should be small.

It is a short calculation (see Appendix J of~\cite{Mayer_organized}) to show that the stable fixed point $N_d^*$ of~\cref{eq:competition} gives the probability distribution $P_d^* = N_d^*/(\sum_{d'} N_{d'}^*)$ for the harm to be minimized. For this to happen, we need
\(
    \varphi(\tilde N_a) = -b' \dd{\bar{F}}/\dd{m} \mid_{m=\rbr{\tilde N_a/N_{\text{st}}}},
\)
where $\tilde N_a = \sum_d N_d^* f_{d,a}$ and $N_{\text{st}}$ is the total number of defenders at the fixed point. The fixed point is found by setting $\dot N_d = 0$. For the nontrivial solution where $N_d > 0$, notice that substituting $\varphi(\tilde{N}_a)$ into~\cref{eq:competition} gives
\(
    \sum_a Q_a \Bar{F}'(\Tilde{P}_a)f_{d,a} = -b'c
\)
which is equivalent to the optimality condition for~\cref{eq:harm}.
% \ch{i was thinking we can just briefly mention the estimation here}

\begin{figure} [htpb]
    \centering
    \includegraphics[width=0.44\linewidth]{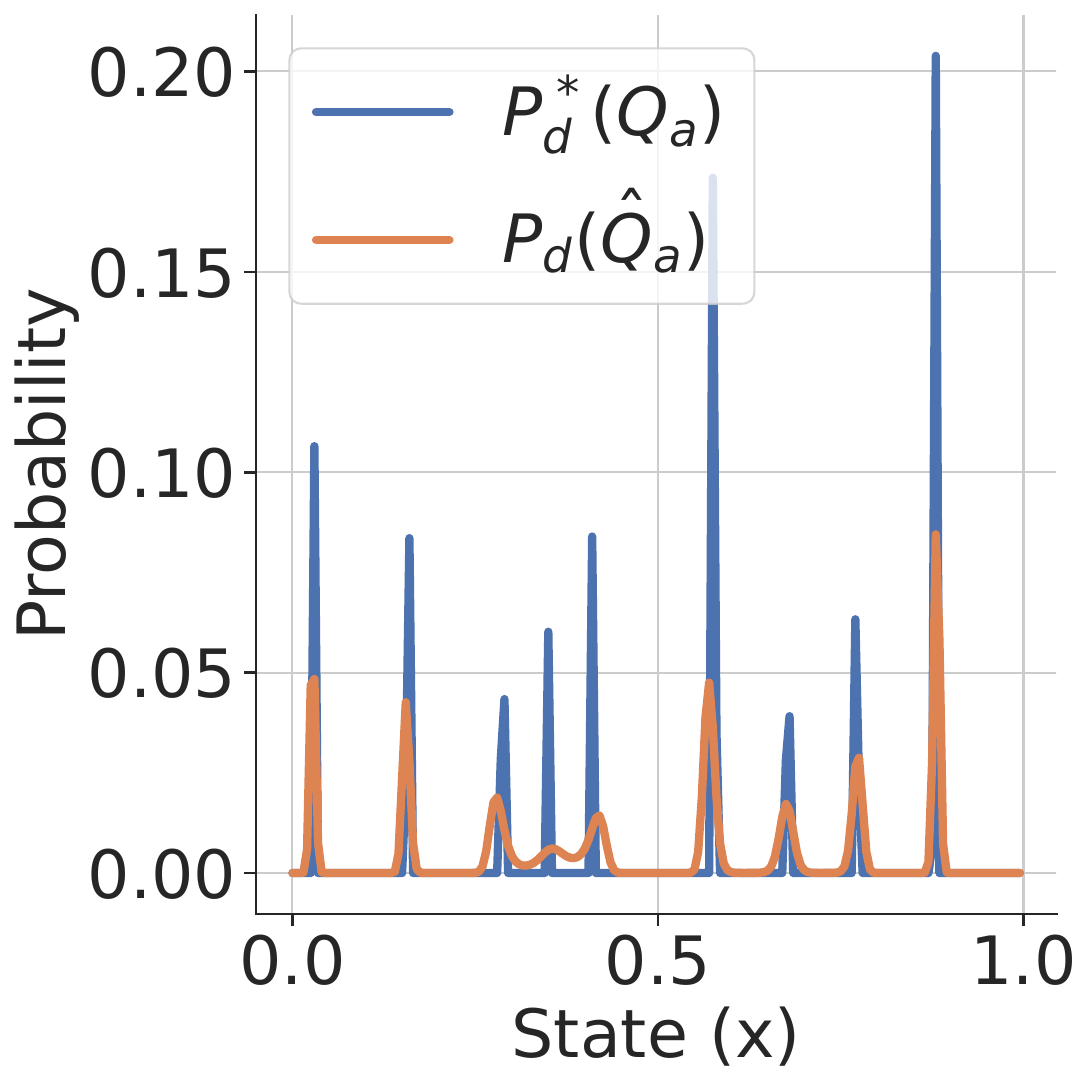}
    \includegraphics[width=0.44\linewidth]{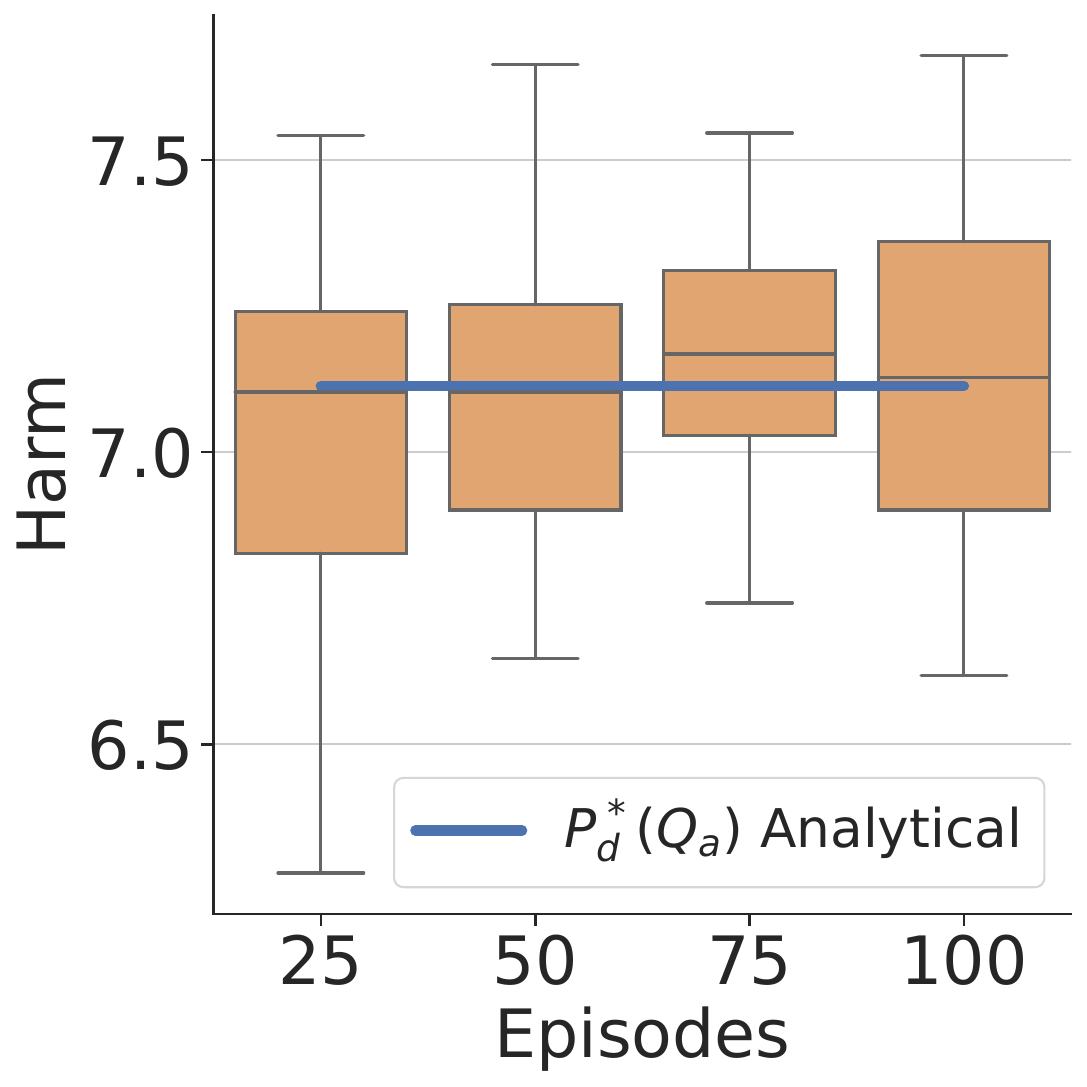}
    \caption{\textbf{Convergence to near-optimal harm with competition dynamics.}
    % For the same setting as that of~\cref{fig:kf_stationary}, we run the dynamics in~\cref{eq:competition} using the estimate $\hat{Q}_a(t)$ calculated using the Kalman filter to induce the defender distribution $P_d(t)$ (instead of calculating it by minimizing the harm in~\cref{eq:harm} as done in~\cref{fig:kf_stationary}).
    % The harm in this case is slightly lower than that in~\cref{fig:kf_stationary}, perhaps because of the transient dynamics in~\cref{eq:competition} which does not lead to $P_d^*$ immediately but models the growth of defenders over time.
    We run the population dynamics in~\cref{eq:competition} to calculate the defender distribution $P_d(t)$ starting from a uniform $P_d(0)$. On the left, we compare the optimal defender distribution (blue) calculated using~\cref{eq:harm} for a known $Q_a$ ($Q_a$ is the same log-normal distribution sampled in~\cref{fig:case0}) with the defender distribution calculated using the competition dynamics (blue) and an estimated $\hat Q_a$ from attacker-defender interactions.
    % For the competition dynamics, the attacker types $\hat Q_a$ are estimated only using attacker-defender interactions as $\hat Q_a(t) = m_a(t)/\sum_a m_a(t)$. It is therefore interesting that the competition dynamics defender team composition matches the optimal one closely. 
    On the right, we show how the empirical harm (orange batched boxplot) incurred by the competition dynamics distribution $P_d(\hat{Q}_a)$ converges towards the analytical harm (blue) and standard deviation shrinks as time progresses. For this experiment, the dynamics were run for $10^4$ iterations per episode, with time in between interactions $\D t = 0.2c^{-1}$, decommission rate $c=0.001$, cross-reactivity bandwidth $\sigma = 0.05$, and $N=200$ types in the shape space.
    }
    \label{fig:competition_stationary}
\end{figure}
% \ch{In Fig. 4 right, can you average the harm for buckets of episodes, say 25, 50, 75, 100 and draw a boxplot? from this picture it does look super convincing that the harm converges. we expect to see the analytical harm come inside the whiskers of the box plot. and mention that it comes inside the whiskers in the caption.}
\cref{fig:competition_stationary} shows $P_d$ obtained using competition dynamics~\cref{eq:competition} and the optimal $P_d^*$ found using~\cref{eq:harm}. In~\cref{fig:competition_stationary} (left), we show the final $P_d$ using competition dynamics after a total $10^6$ interactions ($10^4$ per episode); this simulation uses the same $Q_a$ as~\cref{fig:case0}). In the spirit of adaptation and decentralization, the defender population does not observe the distribution $Q_a$ directly but rather different defenders observe the attackers only via attacker-defender interactions. In our model, this corresponds to the variable $m_a(t)$. The defenders use a simplistic estimator for the attacker distribution: $\hat{Q}_a(t) = m_a(t)/ \sum_a m_a(t)$. This estimate is shared among all the defender types. We see that, over time, the composition of the defender team using competition dynamics converges toward the optimal distribution, and thereby optimal harm even when there are a finite numbers of agents in a discrete shape space.

%% file: RL.tex
\section{Comparison with reinforcement learned policies}
\label{s:RL}

We next use a reinforcement learning (RL) algorithm called soft actor-critic (SAC)~\cite{haarnoja2018soft} to learn a policy for the defender team that minimizes the empirical harm~\cref{eq:Fbar}. Our goal in doing so is to compare the defender team's composition obtained from RL to the one obtained from competition dynamics in~\cref{s:competition}, and to the optimal defender distribution.

%Probabilities $Q_a$ are drawn from a log-normal distribution with coefficient of variation $\k^2 = \exp(\s_Q^2) - 1$ and normalized.
% At each timestep the $N_a$ attackers move synchronously towards the perimeter with constant speed $\|u\|=1$. 

At the beginning of each episode, we initialize with a uniform defender distribution $P_d$ and sample $N_d$ defenders from it. We sample $N_a$ attackers from a $Q_a$, which are fixed across episodes. At each time-step, defenders optimize the parameters $\th$ of a neural network to learn a control policy which outputs a discrete control $[-\D, 0, \D] \ni \bar u \sim \pi_{\theta}(d, \hat Q_a)$ that reallocates a defender agent from type $d$ to type $d+\bar u$; we set $\D = 0.005$. Weights $\th$ are shared by all defender agents. This way, using reinforcement learning, the defender team reallocates its resources using information from the interactions (observations $m_a(t)$ provided to each agent are the same as the competition dynamics formulation, the agents also accumulate them using $\hat{Q}_a(t) = m_a(t)/\sum_a m_a(t)$). During rollouts, at each time-step, attackers proliferate at rate $\nu_a m_a(t)$ and interactions increase with rate is $\nu_a' \l_a(t)$). The reward during a rollout is set to $r = - \sum_a \dot m_a(t)$ which is the rate of proliferation of the attackers; note that the number of attackers that are not recognized grows as $\dot m_a(t) = \nu_a m_a$ so maximizing the reward encourages the policy to recognize attackers. The RL policy should be compared to the competition dynamics in~\cref{s:competition} because defenders compete for interactions with the attackers to earn the reward of successful recognition. Note that at the end of the episode, we can also calculate the effective defender distribution obtained by the SAC policy $P_d = N_d/\sum_{d'} N_{d'}$.

\begin{figure}[!htpb]
    \centering
    \includegraphics[width=0.44\linewidth]{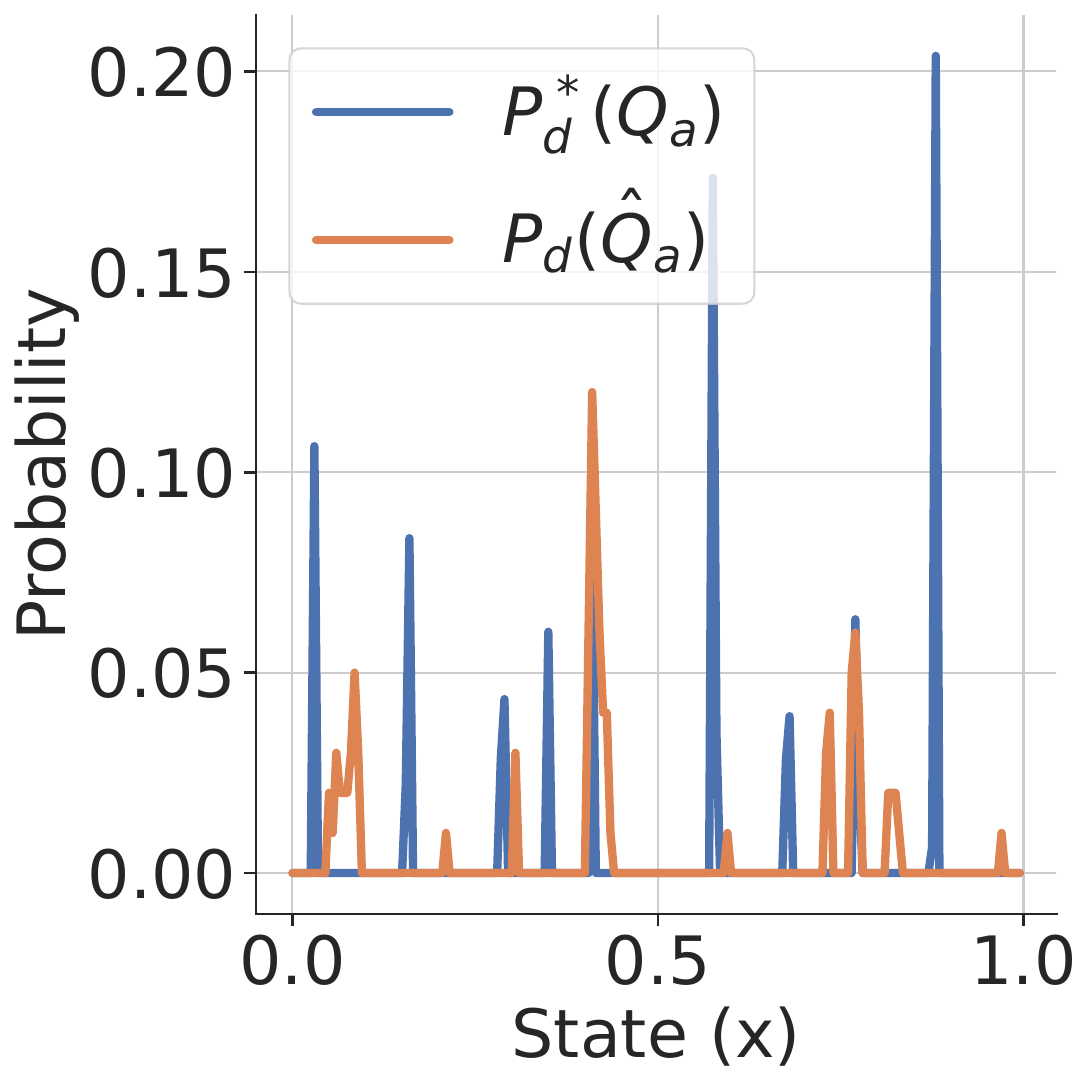}
    \includegraphics[width=0.44\linewidth]{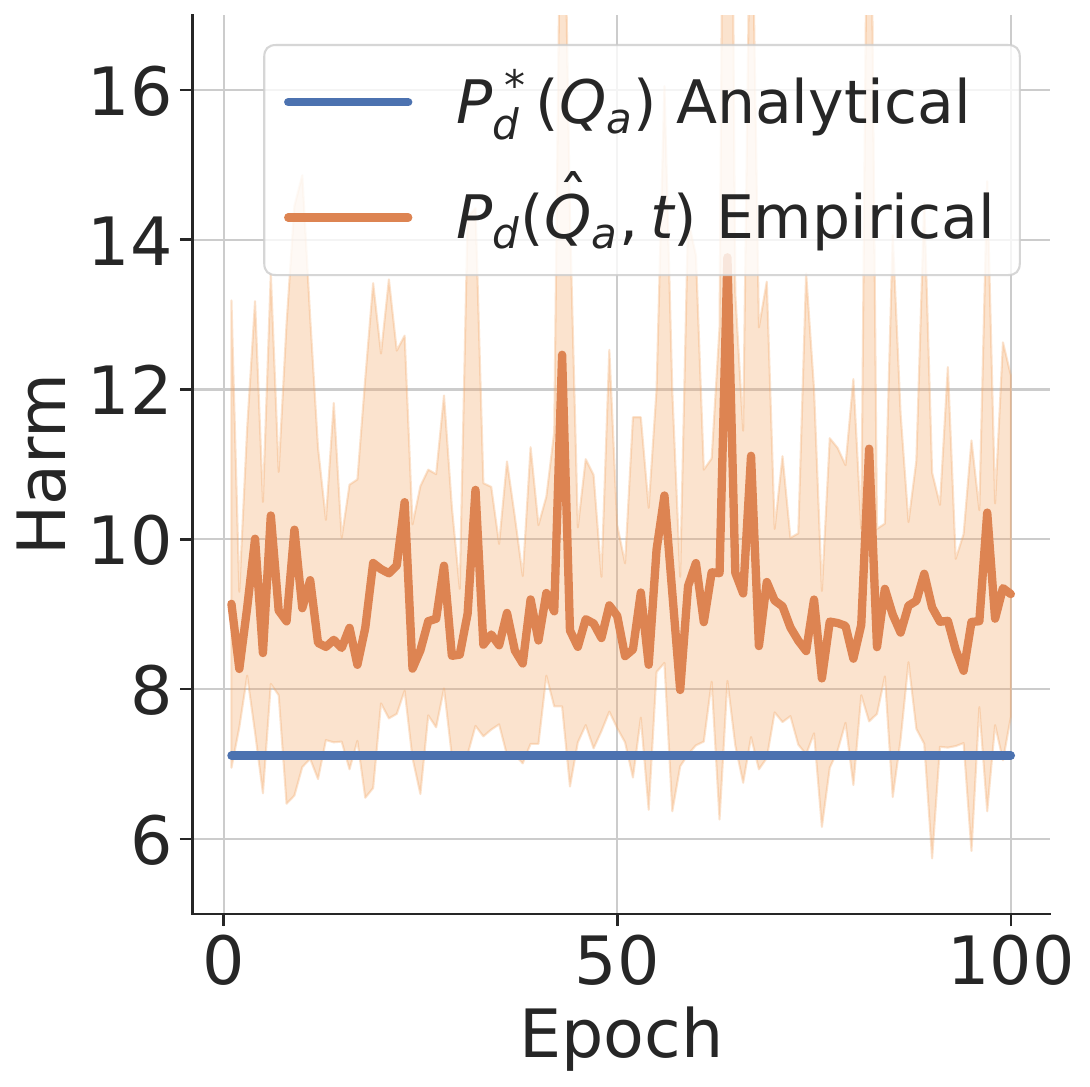}
    \caption{Defender distribution $P_d$ learned by SAC at the end of episode after competing for interactions with attackers sampled from $Q_a$ (the optimal $P_d^*$ is in blue and the same $Q_a$ as~\cref{fig:case0}). We sampled $\sum_d N_d=100$ agents from a uniform distribution that shift states to perform recognition. On the right we compare the test harm of the defender distribution $P_d$ learned by SAC over training epochs to the optimal harm ($10^4$ iterations per epoch for a total of $10^6$ interactions). Cross reactivity bandwidth is $\sigma = 0.05$ and there are $N=200$ types.}
    \label{fig:RL}
\end{figure}

In~\cref{fig:RL}, starting from a uniform distribution $P_d$, we see that the SAC policy changes the defender types such that they collapse to specific types. The distribution obtained by SAC is not the same as the optimal distribution $P_d^*$ and it achieves higher harm (about 9) than the one learned from the competition dynamics in~\cref{fig:competition_stationary} which converges to optimal (about 7). This section shows that: (a) a reinforcement learning formulation for multi-agent interaction problems gives defender team compositions that are close to the optimal ones, and therefore (b) our model can be used as a guiding principle for new RL formulations in situations where abstraction of certain aspects of the problem is possible (e.g., in our case, we have abstracted away the number of agents).

%% file: experiments.tex
% !TEX root = ./root.tex

\section{Mapping the shape space to Euclidean space}
\label{s:mapping_to_euclidean}
% The main change to these type of problems is the determinsitic nature of the following problems. In the immune system, even if the probability of the defender is low, there is an infinite number of defenders so the probability of this interaction is non-zero. In the following problems, there is a finite number of defenders so if 
% Up until now, the number of defenders $N_d$ has not been specified as the theory assumes there will always be an infinite number of defenders to sample from $P_d$. In the following problems, $N_d$ is finite and allocating what types these finite number of defenders is important.
We next discuss how our model can be used to formulate classical multi-agent problems such as perimeter defense~\cite{Shishika2018} and coverage control~\cite{cortes2004} in a very natural fashion. The key idea explored in this section is that we can also think of the shape space as the Euclidean space; defender types $d$ and attacker types $a$ that were rather abstract so far will be likened to the locations of the agents.

% 1D problem
% Explain the cross reactivity as the capture region
\subsection{Perimeter Defense}
% In this example we use a cross-reactivity bandwidth of $\sigma=0.05$ in a shape space $x\in[0,1]$ with discretization width $D=0.005$.
The premise of this game is to capture attacker agents before they breach a perimeter. In literature, capturing attacker agent is usually defined to be point-wise capture in which a defender agent $d$ must be close to an attacker $a$ when $a$ reaches the perimeter, e.g., $\abs{d-a} \leq \epsilon$. In our model, defenders recognize (capture) attackers using the cross-reactivity kernel, e.g., a Gaussian $e^{-(d-a)^2/2\sigma^2}$ where $a$ and $d$ are, say, angles in a polar coordinate frame. Our cross-reactivity can be thought of a capture radius in the classical perimeter-defense literature, except that in our case the probability of capture decreases exponentially as $\abs{d-a}$ increases. Using a small $\sigma$ is akin to point-wise capture.

\begin{figure}[!htpb]
    \centering
    \includegraphics[width=0.42\linewidth]{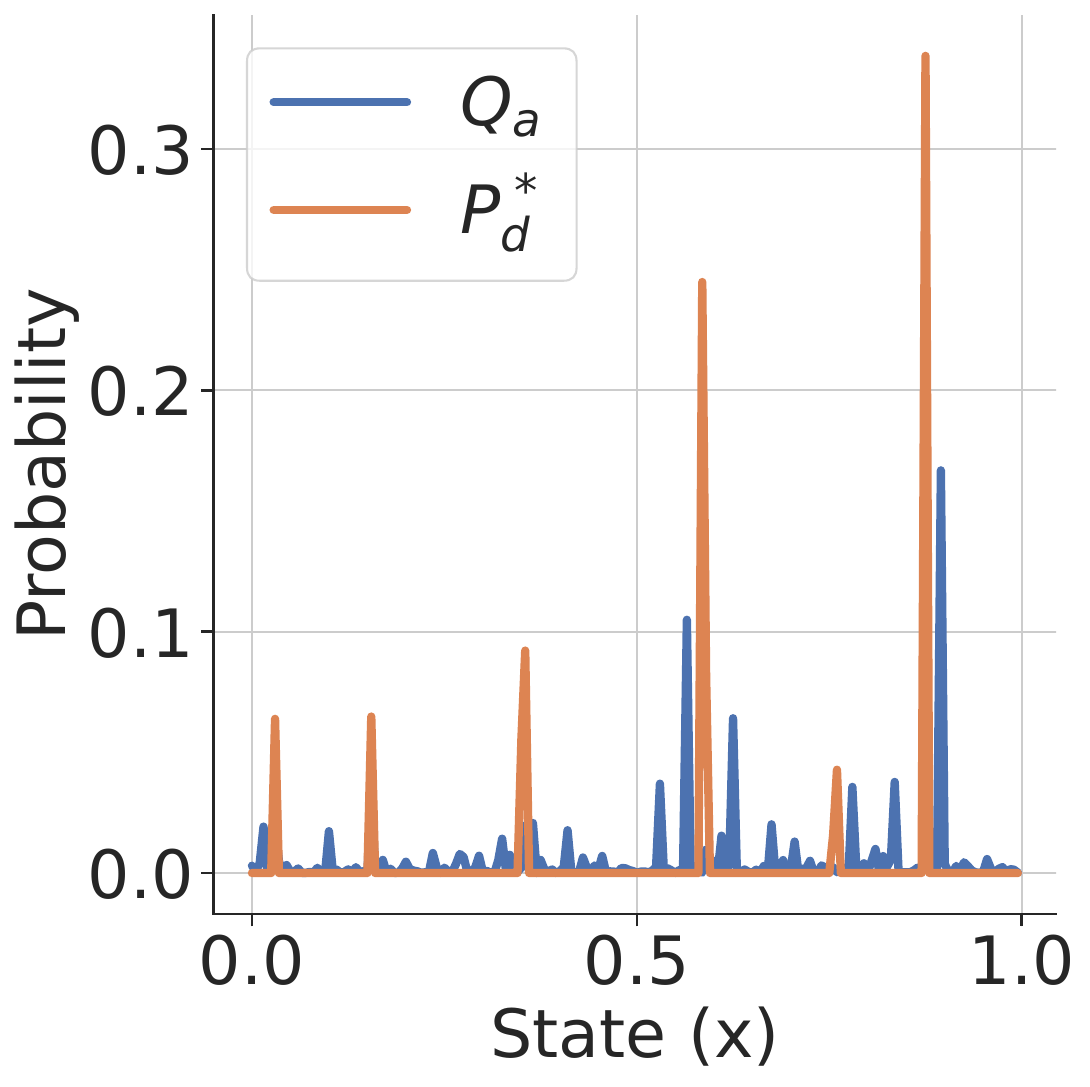}%\\
    \includegraphics[width=0.42\linewidth]{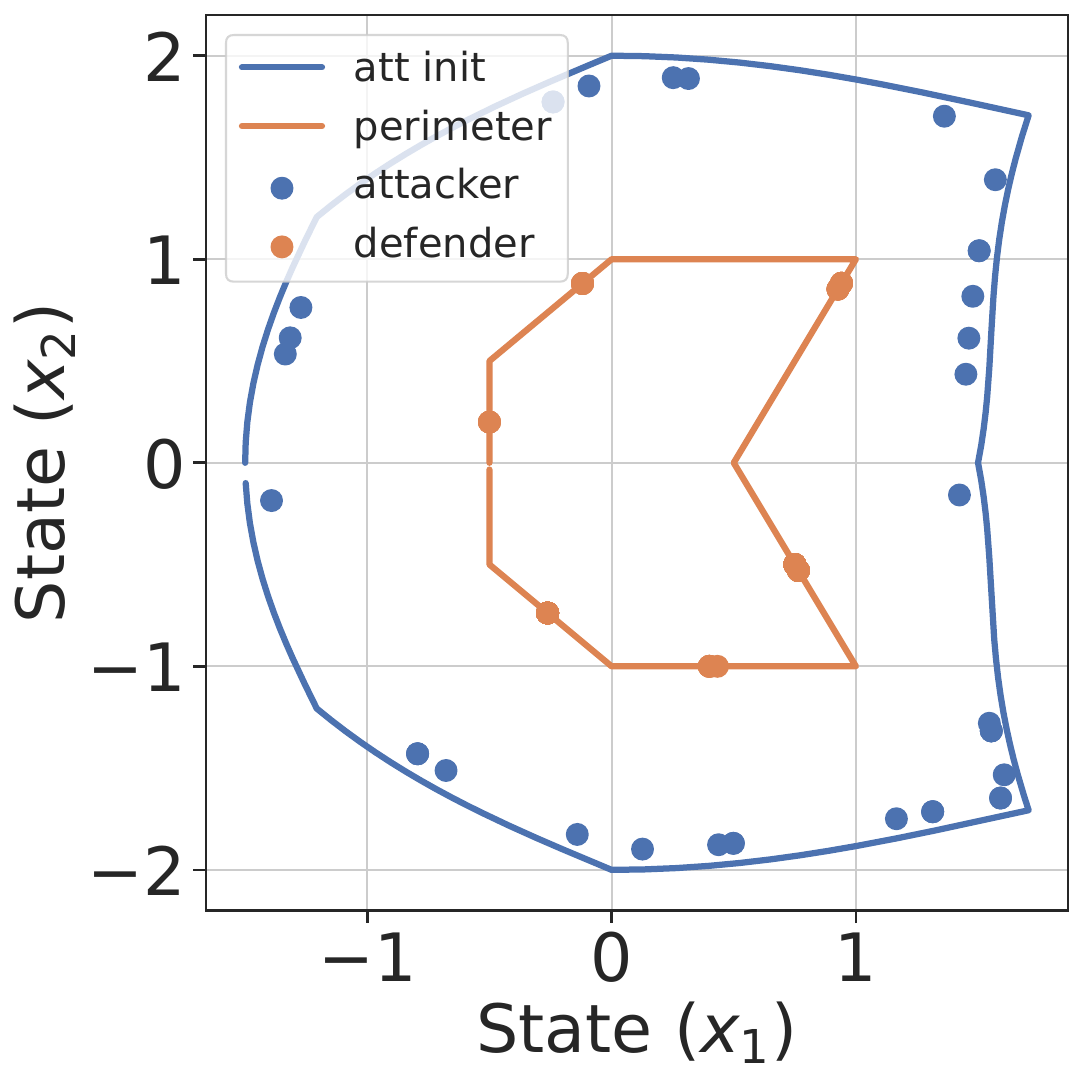}%
    \caption{\textbf{Perimeter Defense Problem.} \textbf{Left:} The standard interaction game and solution where defenders use a cost function $F_a(m) = 1 - e^{-\beta m}$ where $\beta=0.2$, cross-reactivity bandwidth $\s=0.05$ for shape space $x\in[0,1]$ and $N=200$ types.
    \textbf{Right:} We mapped the shape space types shown on the left to polar coordinate angles $x\in [-\pi,\pi]$ and then to the Euclidean space. 100 attackers (blue dots) sampled from $Q_a$ are initialized on the blue line which is 1 polar radius away from the perimeter (orange line). 100 defenders sampled from $P_d^*$ live on the perimeter. At each timestep, attackers move and interactions occur until the end of the episode at time $T$.
    }
    \label{fig:pd}
\end{figure}

In~\cref{fig:pd} (right), defenders (orange) are initialized on the perimeter by sampling from the solution $P_d^*$. Attackers (blue) are initialized at a distance of 1 from this perimeter and move along a fixed polar angle, i.e., $a = (r,\theta)$, at a rate $\Dot{r}$ towards the perimeter. At each time-step, defenders interact with attackers with a Poisson rate $\l_a(t)$ that grows exponentially with each failed interaction. This is our way of modeling the probability of capture based on the distance of the attacker from the perimeter; capture occurs using the cross-reactivity kernel. The game ends after a fixed time $T=r/\Dot{r}$.

As seen by the description of the perimeter defense game, there is little to no difference between a late response and no response if the attacker reaches the perimeter. Therefore, in our model, we use the cost $F_a(m) = 1-e^{-\beta m}$ which saturates at large numbers of interactions $m$. The associated average cost is given as $\Bar{F}_a = \beta/(\beta + \Tilde{P}_a)$. It can be seen that the optimal defender distribution with this type of cost function will forgo parts of the attacker distribution $Q_a$ that have small probability mass, i.e., those attackers will be recognized late. We measure the $\text{capture ratio} = \text{\# captured}/N_a$ in a fixed time horizon $T = 20$ where $T$ is defined by the velocity of the attacker $\Dot{r}=0.05$. We set $\beta = 0.2$ such that the cost function $1-e^{-\beta m}$ saturates around $m=20$.

In~\cref{fig:pd} (left) we show the solution to the optimization problem~\cref{eq:Fbar*} in the shape space form where we optimize the cost function as stated above, $F_a(m) = 1-e^{-\beta m}$, to get the optimal defender distribution. $N_a=100$ attackers are sampled from a log-normal as stated in past sections and $N_d=100$ defenders are sampled from this solution. The analytical harm calculated in this scenario is given to be $\text{Harm}(P_d) = 0.556$. In~\cref{fig:pd} (right) a snapshot of the instantiation of this problem in the euclidean space where the perimeter is a polygon and the shape space $x\in[0,1]$ is first mapped to polar angles $x\in[-\pi,\pi]$ and then to the 2D Euclidean space.
% Converting to euclidean space amounts to multiplying these values by the length of the perimeter and we can think of the types as locations.

We consider two cases in~\cref{tab:perimdef}: (1) when attacker-defender task allocation is \emph{random} or \emph{greedy}, and (2) when defenders are \emph{fixed} in their location or \emph{move} according to a control law $\Dot{d} = -k(d-a)$ for some gain $k > 0$. Conceptually, $k$ should be proportional to the ratio of the velocities of the attackers to that of the defenders.

\begin{table}[!htpb]
    \centering
        \resizebox{\linewidth}{!}{
        \renewcommand{\arraystretch}{1.25}
        \begin{tabular}{l c r r}
        \toprule
        Assignment & Mobility & Capture Ratio & Empirical Harm \\
        \midrule
        Random & $\xmark$ & 0.883 $\pm$ 0.035 & 0.583 $\pm$ 0.032\\
        Random & $\cmark$ & 0.905 $\pm$ 0.035 & 0.565 $\pm$ 0.026\\
        Greedy & $\xmark$ & 0.851 $\pm$ 0.095 & 0.379 $\pm$ 0.078\\
        Greedy & $\cmark$ & 0.942 $\pm$ 0.071 & 0.353 $\pm$ 0.077 \\
        \bottomrule
        \end{tabular}
        }
    \caption{Mean and standard deviation of the capture ratio and empirical harm for different scenarios: \emph{random} verses \emph{greedy} interaction assignment and when defenders are \emph{fixed} at initialization verses when they can \emph{move}.}
    \label{tab:perimdef}
\end{table}

In~\cref{tab:perimdef}, the scenario ``random and no mobility'' is exactly our original model where attackers interact at rate $\l_a(t)$ with a randomly assigned fixed defender. The empirical harm aligns with the analytical harm. Incorporating mobility improves the capture ratio and subsequently the empirical harm. Even greedy attacker-defender assignment improves upon the base scenario. We see that although ``greedy and no mobility'' improves empirical harm, i.e., on average, defenders capture attackers quickly, this allows infrequent attackers to breach the perimeter resulting in a slightly smaller capture ratio. ``Greedy with mobility'' allows defenders to capture even those infrequent attackers; the capture ratio is the largest and empirical harm is the smallest.

\subsection{Coverage Control}
% main change to the model is adjusting the cross reactivity

Coverage control seeks to place multiple sensors at optimal locations in the environment to maximize the information obtained from them. In the context of a fixed environment and a fixed phenomenon being observed, say a function $\phi: Q \to \reals_+$ over the domain $Q$ (say a convex polytope), there are a number of works~\cite{cortes2004} that use Voronoi partitions to establish ``regions of dominance'' and move sensors/agents toward the centroids of the partitions. Suppose the sensing capability degrades with distance $\norm{q-p}^2$. The locations of the $n$ sensors $P=(p_1,\dots, p_n)$ can be chosen to maximize $\sum_{i=1}^n\int_{V_i}\norm{q-p_i}^2\phi(q)dq$ where $V_i = \{ q\in Q \mid \norm{q-p_i} \leq \norm{q-p_j}, \forall j \neq i \}$ is the Voronoi cell.

% at a point $q$ taken from the $i$th sensor at the position $p_i$ degrades with the distance $\norm{q-p_i}^2$ between $q$ and $p_i$. A partition, also known as a Voronoi cell, of $Q$ is a collection of polytopes $V = \{V_i,\dots,V_n\}$ with disjoint interiors whose union is $Q$. The $i$th agent at the position $p_i$ is responsible for measurements over its Voronoi cell $V_i = \{q\in Q| \norm{q-p_i}\leq \norm{q-p_i}, \forall j \neq i\}$. The locational optimization problem is given as $\mathcal{H}(P,V) = \sum_{i=1}^n\int_{V_i}\norm{q-p_i}^2\phi(q)dq$.

% The local optimal solution is given as $x_i^* = \arg\min H(p_i,\dots,p_n) = \frac{\int_{V_i}q \phi(q)dq}{\int_{V_i}\phi(q)dq} = C_{V_i}$ where $C_{V_i}$ is the centroid of each Voronoi cell $V_i$. They often will then use a decentralized gradient based move-to-centroid controller $\Dot{p}_i = k(C_{V_i} - p_i)$ where $k$ is some user-defined gain.

\begin{figure}
\centering
\includegraphics[width=0.32\linewidth]{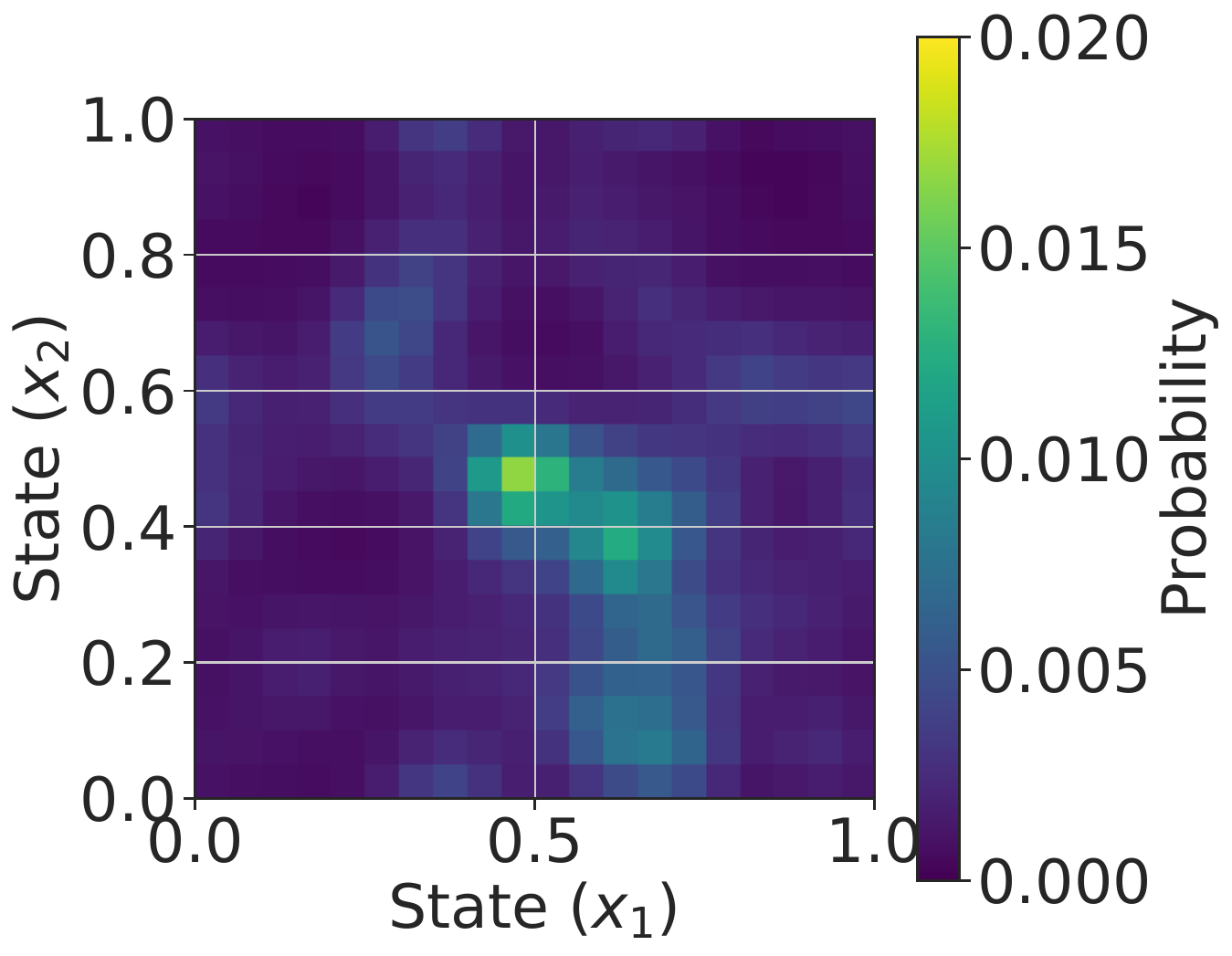}
\includegraphics[width=0.32\linewidth]{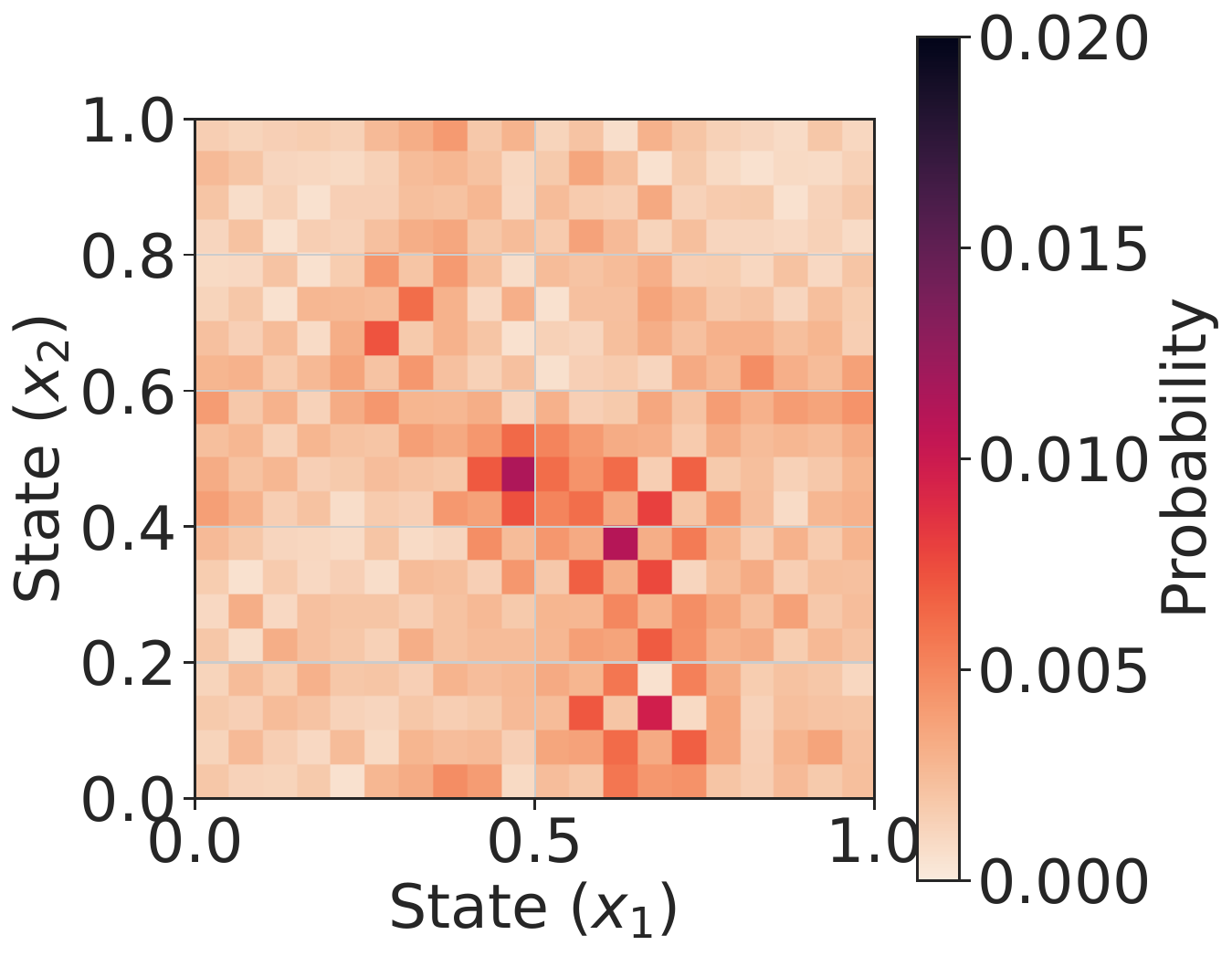}
\includegraphics[width=0.32\linewidth]{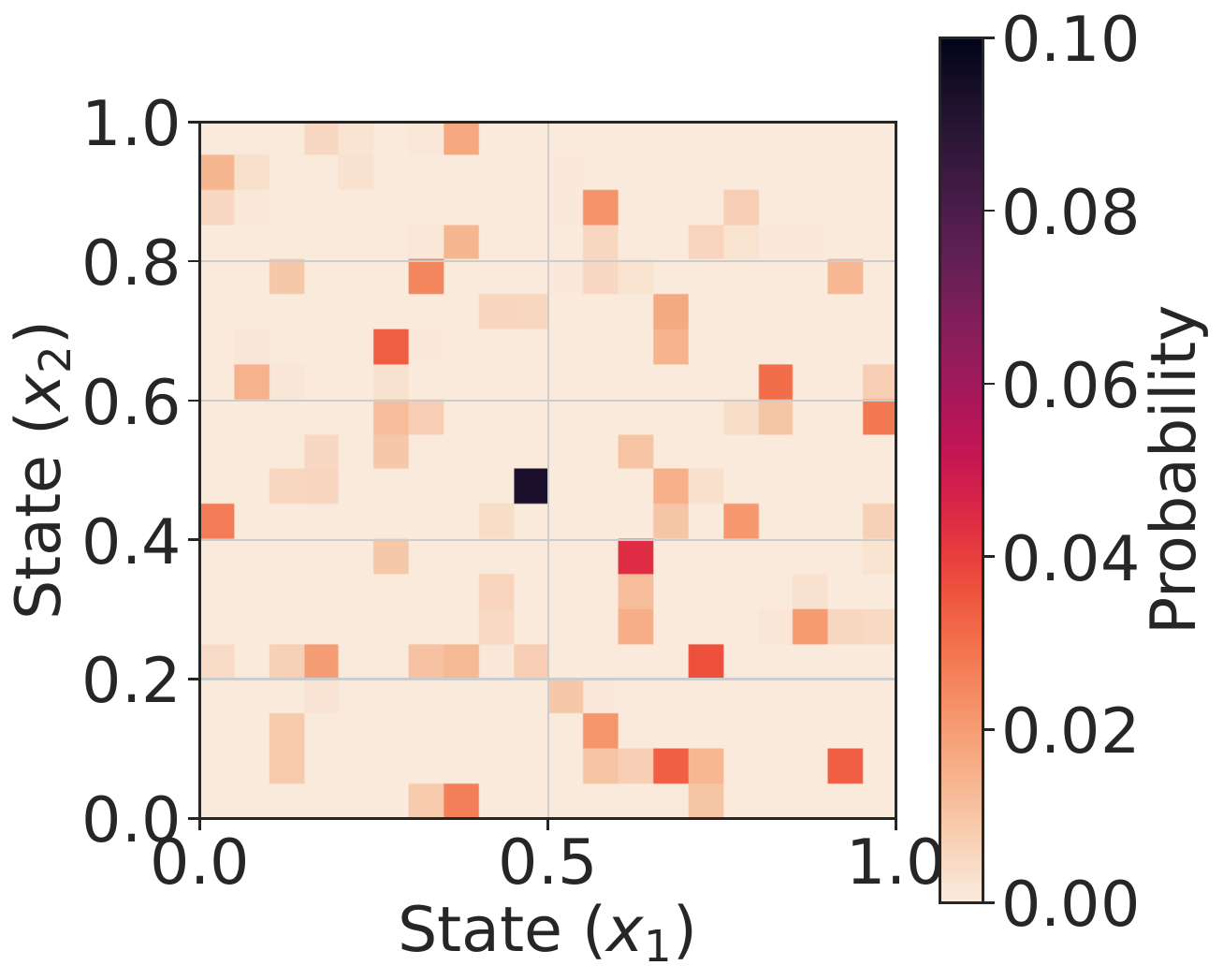}
\caption{In the coverage control problem, we model the shape space as the two-dimensional Euclidean space where the state $(x_1, x_2) \in [0,1]^2$ and $N=400$ discrete locations. Given a distribution $\phi \equiv Q_a \tilde F_a$ with $\sigma_Q = 0.05$ and $F_a(m) = m$, (blue, left) we show the optimal $P_d^*$ for two different cross-reactivity bandwidths $\sigma_P = 0.04$ (orange, middle) and $\sigma_P = 0.08$ (orange, right).
% \textbf{Left:} $Q_a$ with $\sigma_Q = 0.05$
% \textbf{Middle:} $P_d$ with $\sigma_P = 0.04$
% \textbf{Right:} $P_d$ with $\sigma_P = 0.08$
}
\label{fig:cov_control}
\end{figure}

Our model maps naturally to such settings, the coverage objective is akin to the harm, i.e., $\phi \equiv Q_a \bar F_a$. The sensing degradation model $\norm{q-p}^2$ corresponds to the cross-reactivity kernel. But while the coverage control formulation first selects a fixed number of sensors $n$, our formulation only optimizes the locations of an arbitrary number of sensors (replace the summation by an integral in~\cref{eq:numerical_harm}). This is a major difference in the two approaches and leads to a very interesting phenomenon shown in~\cref{fig:cov_control}. For a small chosen $\s$, the defenders cover essentially the high-probability regions of $\phi$; this solution would be similar to the one obtained by, say, Voronoi partition-based methods. For a large $\s$, the sensors are much more spread out in the domain. There are some sensors in extremely low probability regions (but due to their cross-reactivity then sense large parts of the domain). This suggests that our model is well-suited to addressing coverage control problems where the number of sensors needs to be \emph{minimized} (as opposed to simply having a preselected number of sensors).

% Coverage control dictates distinct and disjoint partitions with which a robot lives in. The partitions do not have a specific shape and are dependent on its neighbors. If $Q$ were a single Gaussian with a single mode. Two agents in the coverage control formulation will balance their locations such that partition the space equally.
% In comparison, a Gaussian dominance region does not require disjoint coverage but rather coverage comes implicitly. In the same scenario above, our formulation will produce a solution in which depending on the cross reactivity width, will focus its mass on the mode of the Gaussian.

% where we have a similar formulation but we can reformulate our problem if our cross-reactivity function is given as $\frac{1}{(d-a)^2}$.

% [cortes 2004]. Decentralized, adaptive coverage control for networked robots [schwager 2008][wenhao, brian reilly]. they move to minimize a cost function representing the collective sensing cost of the network. Introduces a notion of sensor coverage that formalizes an optimal sensor placement problem  utilizing lloyd algorithm

% Similar to our description of the perimeter defense problem we have differences than the coverage control problem which would be 2D instantiation of our problem. In our setting rather than there being a deterministic location in which the points of interest are and there is a signal to noise ration which degrades with distance, we model a stochastic problem where there is a probability that the attacker might be in a euclidean location. 

% Th we have an integral over the cross reactivity of the agent locations whereas they have an integral over the voronoi regions. 
% The main difference between our objective functions is that we have a probability of interaction even though attacker $a$ and defender $d$ has a distance $\norm{a-d}^2 \leq \epsilon$ they do not necessarily interact at each time step and therefore we are solving similar but fundamentally different problems. 

% However, we can still showcase what our model's solution would be in the 2D case. 

%% file: conclusion.tex
% !TEX root = ./root.tex

\section{Conclusion}
\label{s:conclusion}

We used a mathematical model to understand what an optimal defender team composition should be. The key property of this model is the cross-reactivity which enables defender agents of a given type to recognize attackers, of a few different types. This allows the defender distribution to be supported on a discrete set, even if the shape-space, i.e., the number of different types of agents, is very large. Cross-reactivity is also fundamentally responsible for the defender team to be able to estimate an unknown and evolving attacker distribution. This model points to a number of guiding principles for the design of multi-agent systems across many different scales and problem settings, e.g., the immune system where the population-level dynamics of the attackers and defenders that was used to formulate the model, to various experimental settings where we evaluated the model with a finite number of agents, to competition dynamics that allows effective decentralized control policies and reinforcement learning such controllers.

As discussed in~\cref{s:competition}, the attacking distribution might not always be known. For stationary distributions, the naive version presented works well. However, we have also implemented a Kalman filter to obtain a better estimate of the attacker team's composition. This estimate is used to update the defender team's composition at the end of each episode. Due to space limitations, we do not elaborate upon these results, but the Kalman filter can also effectively address situations where the attacker team's composition changes over time.

% We have also obtained similar results as those above, for the case when defenders move and tackle a stationary attacker distribution from which agents are sampled (see~\cref{s:mapping_to_euclidean} for how we implement this, the Kalman filter in this case estimates the locations of the attackers). 

%In the future, we hope to seek parallels of these ideas with specific strategies and the evolution of these strategies in multi-agent games such as soccer or basketball. There are certain similarities: teams like to be constituted of players that can cover the opponent team's repertoire of speed, skill, size and strength. This is an example of ``zone'' defense and a generalist strategy; it prevents a specific opponent from maximizing their capabilities. Man-to-man strategies, which would be akin to a small cross-reactivity bandwidth in our model, work well when the opponent distribution is known well, but they are not robust to changes or incorrect estimates of the opponent distribution. The key difference in these problems to those explored here is that these problems are highly dynamic games with evolving strategies, objectives, and consequently dynamic equilibria.

%% file: root.bbl
\begin{thebibliography}{10}

\bibitem{Mayer_organized}
A.~Mayer, V.~Balasubramanian, T.~Mora, and A.~M. Walczak, ``How a well-adapted
  immune system is organized,'' {\em PNAS}, vol.~112, no.~19, pp.~5950--5955,
  2015.

\bibitem{Shishika2018}
D.~Shishika and V.~Kumar, ``Local-game decomposition for multiplayer
  perimeter-defense problem,'' pp.~2093--2100, 12 2018.

\bibitem{cortes2004}
J.~Cortes, S.~Martinez, T.~Karatas, and F.~Bullo, ``Coverage control for mobile
  sensing networks,'' {\em IEEE Transactions on Robotics and Automation},
  vol.~20, no.~2, pp.~243--255, 2004.

\bibitem{Karaman2010}
S.~Karaman and E.~Frazzoli, ``Incremental sampling-based algorithms for a class
  of pursuit-evasion games,'' vol.~68, pp.~71--87, 01 2010.

\bibitem{isaacs1954}
R.~Isaacs, {\em Differential Games I}.
\newblock RAND Corporation, 1954.

\bibitem{Lee-GroundvsAir}
E.~S. Lee, D.~Shishika, G.~Loianno, and V.~Kumar, ``Defending a perimeter from
  a ground intruder using an aerial defender: Theory and practice,'' in {\em
  SSRR}, pp.~184--189, 2021.

\bibitem{Chen-ReachAvoid}
M.~Chen, Z.~Zhou, and C.~J. Tomlin, ``A path defense approach to the
  multiplayer reach-avoid game,'' in {\em CDC}, pp.~2420--2426, 2014.

\bibitem{Schwager2009}
M.~Schwager, D.~Rus, and J.-J. Slotine, ``Decentralized, adaptive coverage
  control for networked robots,'' {\em The International Journal of Robotics
  Research}, vol.~28, no.~3, pp.~357--375, 2009.

\bibitem{luo2018AdaptiveSampling}
W.~Luo and K.~Sycara, ``Adaptive sampling and online learning in multi-robot
  sensor coverage with mixture of gaussian processes,'' in {\em 2018 IEEE
  International Conference on Robotics and Automation (ICRA)}, pp.~6359--6364,
  2018.

\bibitem{Reily2021}
B.~Reily, T.~Mott, and H.~Zhang, ``Adaptation to team composition changes for
  heterogeneous multi-robot sensor coverage,'' in {\em 2021 IEEE International
  Conference on Robotics and Automation (ICRA)}, pp.~9051--9057, 2021.

\bibitem{lowe2020multiagent}
R.~Lowe, Y.~Wu, A.~Tamar, J.~Harb, P.~Abbeel, and I.~Mordatch, ``Multi-agent
  actor-critic for mixed cooperative-competitive environments.'' arXiv
  1706.02275, 2020.

\bibitem{long2020evolutionary}
Q.~Long, Z.~Zhou, A.~Gupta, F.~Fang, Y.~Wu, and X.~Wang, ``Evolutionary
  population curriculum for scaling multi-agent reinforcement learning.'' arXiv
  2003.10423, 2020.

\bibitem{hsu2020scalable}
C.~D. Hsu, H.~Jeong, G.~J. Pappas, and P.~Chaudhari, ``Scalable reinforcement
  learning policies for multi-agent control,'' in {\em IROS}, pp.~4785--4791,
  2020.

\bibitem{Hsieh2008BiologicallyIR}
M.~A. Hsieh, {\'A}.~M. Hal{\'a}sz, S.~Berman, and V.~R. Kumar, ``Biologically
  inspired redistribution of a swarm of robots among multiple sites,'' {\em
  Swarm Intelligence}, vol.~2, pp.~121--141, 2008.

\bibitem{Zavlanos-controlforflocking}
M.~M. Zavlanos, H.~G. Tanner, A.~Jadbabaie, and G.~J. Pappas, ``Hybrid control
  for connectivity preserving flocking,'' {\em TAC}, vol.~54, no.~12,
  pp.~2869--2875, 2009.

\bibitem{Bettini-HetGPPO}
M.~Bettini, A.~Shankar, and A.~Prorok, ``Heterogeneous multi-robot
  reinforcement learning,'' 2023.

\bibitem{Mitra-HeterogeneityBandit}
A.~Mitra, H.~Hassani, and G.~Pappas, ``Exploiting heterogeneity in robust
  federated best-arm identification,'' 2021.

\bibitem{Wang-RODE}
T.~Wang, T.~Gupta, A.~Mahajan, B.~Peng, S.~Whiteson, and C.~Zhang, ``Rode:
  Learning roles to decompose multi-agent tasks.'' arXiv 2010.01523, 2020.

\bibitem{Salam-heterosensors}
T.~Salam and M.~A. Hsieh, ``Heterogeneous robot teams for modeling and
  prediction of multiscale environmental processes.'' arXiv 2103.10383, 2021.

\bibitem{Edwards-heterocontrol}
V.~Edwards, P.~Rezeck, L.~Chaimowicz, and M.~A. Hsieh, ``{Segregation of
  Heterogeneous Robotics Swarms via Convex Optimization},'' Dynamic Systems and
  Control Conference, 10 2016.

\bibitem{Grocholsky-coopairground}
B.~Grocholsky, J.~Keller, V.~Kumar, and G.~Pappas, ``Cooperative air and ground
  surveillance,'' {\em RAM}, vol.~13, no.~3, pp.~16--25, 2006.

\bibitem{TwuEgerstedt2014}
P.~Twu, Y.~Mostofi, and M.~Egerstedt, ``A measure of heterogeneity in
  multi-agent systems,'' in {\em 2014 American Control Conference},
  pp.~3972--3977, 2014.

\bibitem{Abbas2011DistributionOA}
W.~Abbas and M.~Egerstedt, ``Distribution of agents in heterogeneous multiagent
  systems,'' {\em IEEE Conference on Decision and Control and European Control
  Conference}, pp.~976--981, 2011.

\bibitem{Bezzo2013}
N.~Bezzo, R.~A. Cortez, and R.~Fierro, {\em Exploiting Heterogeneity in Robotic
  Networks}, pp.~53--75.
\newblock Berlin, Heidelberg: Springer Berlin Heidelberg, 2013.

\bibitem{chernova-Strata}
H.~Ravichandar, K.~Shaw, and S.~Chernova, ``Strata: A unified framework for
  task assignments in large teams of heterogeneous agents,'' 2019.

\bibitem{PERELSON1979645}
A.~S. Perelson and G.~F. Oster, ``Theoretical studies of clonal selection:
  Minimal antibody repertoire size and reliability of self-non-self
  discrimination,'' {\em Journal of Theoretical Biology}, vol.~81, no.~4,
  pp.~645--670, 1979.

\bibitem{santambrogioOptimalTransportApplied2015}
F.~Santambrogio, ``Optimal transport for applied mathematicians,'' {\em
  Birk\"auser, NY}, vol.~55, no.~58-63, p.~94, 2015.

\bibitem{rockafellar2000optimization}
R.~T. Rockafellar, S.~Uryasev, {\em et~al.}, ``Optimization of conditional
  value-at-risk,'' {\em Journal of risk}, vol.~2, pp.~21--42, 2000.

\bibitem{DEBOER1994375}
R.~J. {De Boer} and A.~S. Perelson, ``T cell repertoires and competitive
  exclusion,'' {\em Journal of Theoretical Biology}, vol.~169, no.~4,
  pp.~375--390, 1994.

\bibitem{haarnoja2018soft}
T.~Haarnoja, A.~Zhou, P.~Abbeel, and S.~Levine, ``Soft actor-critic: Off-policy
  maximum entropy deep reinforcement learning with a stochastic actor,'' 2018.

\end{thebibliography}
